\documentclass[aps,prb,twocolumn,english,showpacs,superscriptaddress,amssymb,amsfonts]{revtex4}
\usepackage[T1]{fontenc}
\usepackage[latin9]{inputenc}
\usepackage{amsmath}
\usepackage{dsfont}
\usepackage{amstext}

\usepackage{tocvsec2}

\usepackage{amssymb}
\usepackage{amsbsy}
\usepackage{amsthm}
\usepackage{epsfig}
\usepackage{framed}
\usepackage{graphicx}
\usepackage{bbm}
\usepackage{hyperref} %turn this back on later but it makes so many errors
\usepackage{color}
\usepackage{multirow}
\usepackage{lscape}

\newcommand{\Ref}[1]{Ref.~\onlinecite{#1}}

\newcommand{\bst}{{\mathcal{T}}}
\newcommand{\bsm}{\mathcal{M}}
\newcommand{\bsi}{\mathcal{I}}
\newcommand{\bss}{\mathcal{S}}
\newcommand{\bsc}{{\mathcal{C}}}

\newcommand{\ie}{{\emph{i.e.~}}}
\makeatletter

\newcommand{\Rmnum}[1]{\expandafter\@slowromancap\romannumeral #1@}
\makeatother
\newcommand{\imth}{\hspace{1pt}\mathrm{i}\hspace{1pt}}
\newcommand{\alert}[1]{{\color{red}{#1}}}
\newcommand{\eg}{{\emph{e.g.~}}}

\newcommand{\mbz}{{\mathbb{Z}}}
\newcommand{\bea}{\begin{eqnarray}}
\newcommand{\eea}{\end{eqnarray}}
\newcommand{\bpm}{\begin{pmatrix}}
\newcommand{\epm}{\end{pmatrix}}
\newcommand{\bal}{\begin{aligned}}
\newcommand{\eal}{\end{aligned}}

\newcommand{\dket}[1]{|{#1}\rangle}

\newcommand{\gpc}[3]{{\mathcal{H}^{#1}\big({#2},{#3}\big)}}

\makeatother

\usepackage{babel}
\begin{document}
\title{Classification and construction of higher-order symmetry protected topological phases of interacting bosons}

\author{Alex Rasmussen}
\author{Yuan-Ming Lu}
\affiliation{Department of Physics, The Ohio State University, Columbus, OH 43210, USA}

\begin{abstract}
Motivated by the recent discovery of higher-order topological insulators, we study their counterparts in strongly interacting bosons: ``higher-order symmetry protected topological (HOSPT) phases''. While the usual (1st-order) SPT phases in $d$ spatial dimensions support anomalous $(d-1)$-dimensional surface states, HOSPT phases in $d$ dimensions are characterized by topological boundary states of dimension $(d-2)$ or smaller, protected by certain global symmetries and robust against disorders. Based on a dimensional reduction analysis, we show that HOSPT phases can be built from lower-dimensional SPT phases in a way that preserves the associated crystalline symmetries. When the total symmetry is a direct product of global and crystalline symmetry groups, we are able to classify the HOSPT phases using the K\"unneth formula of group cohomology. Based on a decorated domain wall picture of the K\"unneth formula, we show how to systematically construct the HOSPT phases, and demonstrate our construction with many examples in two and three dimensions.
\end{abstract}

\pacs{}

\maketitle
%\setcounter{tocdepth}{2}
%\begin{widetext}
%\tableofcontents
%\end{widetext}

%\maxsecnumdepth{subsection}

%{\small \setcounter{tocdepth}{2} \tableofcontents}

%\twocolumn

\section{Introduction}

The discovery of topological insulators\cite{Hasan2010,Hasan2011,Qi2011c} (TIs) unveiled a large class of symmetry protected topological (SPT) states\cite{Chen2013,Senthil2015}, which in $d$ spatial dimensions feature symmetry-protected surface states on $(d-1)$-dimensional open boundaries, such as one-dimensional (1d) helical edge states in two-dimensional (2d) quantum spin Hall insulators\cite{Maciejko2011} and 2d Dirac fermions on the surface of three-dimensional (3d) topological insulators\cite{Hasan2011}. Recently a new family of ``higher-order'' topological insulators has been revealed\cite{Benalcazar2017,Benalcazar2017a,Song2017,Langbehn2017,Fang2017,Khalaf2017,Schindler2018a,Ezawa2018,Yan2018,Wang2018b,Wang2018c,Trifunovic2018,Wang2018a,Matsugatani2018,Franca2018,Serra-Garcia2018,Schindler2018b,Peterson2018,Zhang2018,slager2014,calugaru2018}, which do not have gapless surface states, but exhibit gapless modes on hinges and corners of the system. Generally a $k$-th order TI in $d$ dimensions hosts robust gapless excitations on $(d-k)$-dimensional open boundaries of the system: such as 0-dimensional corner states in 2nd-order 2d TIs and 3rd-order 3d TIs, as well as 1d hinge states in 2nd-order 3d TIs. In this terminology, the usual TIs can be called 1st-order TIs. These lower dimensional boundary excitations are robust against any small perturbations such as disorders and crystal distortions, as long as the global symmetry $G_0$ is protected, analogous to the stability of the TI surface states. It has been shown that the higher-order TIs usually also preserve certain crystalline symmetries in addition to the global symmetry\cite{Benalcazar2017,Benalcazar2017a,Song2017,Langbehn2017,Fang2017,Khalaf2017,Schindler2018a,Trifunovic2018,Song2018}.
While most of the efforts so far are focused on higher-order topological phases within band theory of non-interacting fermions, little is known about their strongly-interacting counterparts in \eg interacting boson systems\cite{Dubinkin2018,You2018}. How to understand the higher-order SPT phases in a generic interacting boson system?

The goal of this work is to address this issue. We provide the classification and explicit construction for ``strong'' higher-order SPT (HOSPT) phases of interacting bosons with various global (\ie onsite) symmetry $G_0$ and crystalline symmetry $G_c$, whose lower-dimensional boundary excitations are protected only by onsite symmetry $G_0$ and hence robust against disorders and crystal distortions. We show that a $k$-th order SPT phase in $d$ spatial dimensions is built by by stacking $(d+1-k)$-dimensional $G_0$-SPT phases, in a way which preserves crystalline symmetry $G_c$. In particular for total symmetry $G=G_0\times G_c$ as a direct product of onsite symmetry $G_0$ and crystalline symmetry $G_c$,, we show that all $(k+1)$-th order SPT phases in $d$ dimensions are classified within the group cohomology
\bea\label{classification}
\gpc{k}{G_c^\ast}{\gpc{d+1-k}{G_0}{U(1)}}.
\eea
where $G_c^\ast$ is isomorphic to crystalline group $G_c$ by regarding each orientation-reversing symmetry operation as an anti-unitary operator\cite{Jiang2017,Thorngren2018}. The above classification also provides a procedure to construct these HOSPT phases from building blocks of $(d-k)$-dimensional SPT phases protected by onsite symmetry $G_0$ only, as illustrated in many examples.

This work is organized as follows. First in section \ref{sec:picture} we discuss the physical picture behind HOSPT phases based on a dimensional reduction point of view. Then we show the general classification of HOSPT phases in section \ref{sec:classification} based on the Kennuth formula of group cohomology, and how to the construct the HOSPT phases using the decorated domain wall picture in section \ref{sec:ddw}. The classification and construction are demonstrated for 2nd-order SPT phases in two (section \ref{sec:k=1,d=2}) and three (section \ref{sec:k=1,d=3}) dimensions, and 3rd-order SPT phases in three dimensions (\ref{sec:k=2,d=3}). We conclude with a few remarks in section \ref{sec:discussions}.

\section{The physical picture}\label{sec:picture}

%\input{section2.tex}
% move into main file when done

Before introducing the mathematical classification for higher-order SPT phases, we first discuss an intuitive physical picture which shows how higher-order SPT phases can be built by stacking lower-dimensional SPT phases. Throughout this work, we will focus on the simplest situation where the total symmetry group $G=G_c\times G_0$ is a direct product of crystalline symmetry group $G_c$ and onsite (\ie global) symmetry group $G_0$.

By definition, a $k$-th order SPT phase in $d$ dimensions is characterized by symmetry protected gapless states on boundaries of $(d-k)$ dimensions. For example, as illustrated in FIG. \ref{fig:C4}, a 2nd-order SPT phases in $d=2$ with 4-fold rotational symmetry $G_c=C_4$ hosts gapless zero modes at each corner of a square-shaped system, which are protected by onsite symmetry $G_0$. Based on this example and without loss of generality, below we present two arguments to establish a dimensional reduction picture for the HOSPT phases: while the 1st argument (section \ref{sec:argument A}) shows why a $k$-th order SPT phase in $d$ dimensions is related to the usual $G_0$-SPT phases in $(d+1-k)$ dimensions, the 2nd argument (section \ref{sec:argument B}) explicitly demonstrates how to build such a HOSPT phase from lower-dimensional SPT phases. While the 1st argument explains why the classification of HOSPT phases is determined by the classification of $(d+1-k)$-dimensional SPT phases, the 2nd argument shows which of the $(d+1-k)$-dimensional SPT phases can consistently lead to a gapped symmetric $k$-th order SPT phases in $d$ dimensions, to be compatible with the crystalline symmetry $G_c$.

\subsection{Corner/hinge states as gapless defects on the gapped open surface}\label{sec:argument A}

\begin{figure}
\includegraphics[width=150pt]{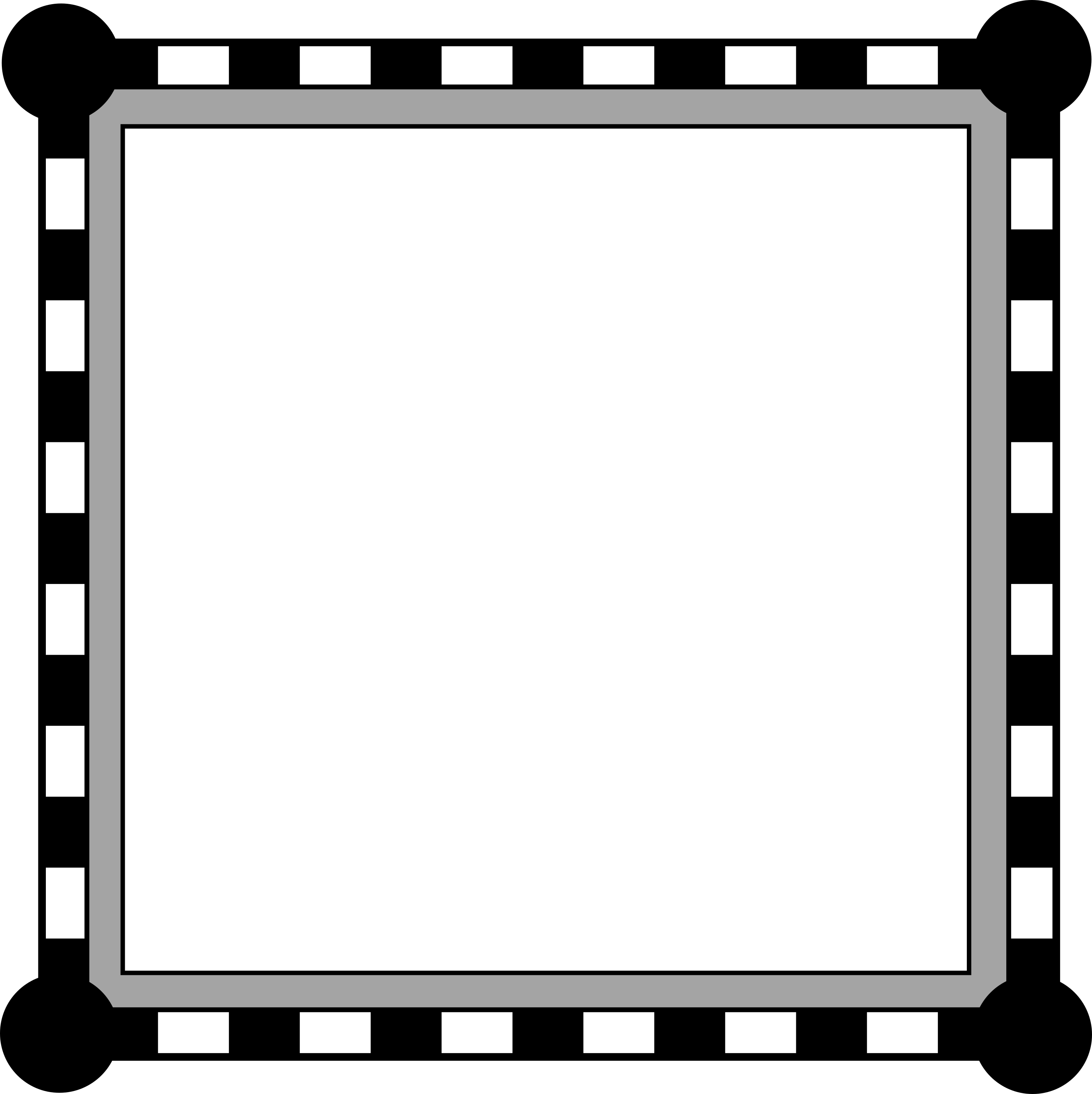}
\caption{Disentangling the gapped bulk with gapless corner states. The interior region $\mathcal{B}$ is colored in white, while the ``cushion'' region $\mathcal{B}_0\setminus\mathcal{B}$ is colored in gray. Finite-depth quantum circuit $\hat U_{\mathcal{B}_0}$ preserving onsite symmetry $G_0$ will trivialize the interior $\mathcal{B}$ into a product state, while keeping the boundary $\bar{\mathcal{B}_0}=\mathcal{A}\setminus\mathcal{B}_0$ (including the gapless corner states) untouched.}\label{fig:disentangle}
\end{figure}

We consider a generic HOSPT phase $\dket{\psi}$ of order $k\geq2$ on a $d$-dimensional open manifold $\mathcal{A}$ (such as the square-shaped system in FIG. \ref{fig:disentangle}), which is gapped almost everywhere except for a $(d-k)$-dimensional submanifold on the boundary $\partial\mathcal{A}$ (such as the four corners with $k=d=2$ in FIG. \ref{fig:disentangle}). Since by definition the $(d-1)$-dimensional boundary $\partial\mathcal{A}$ is gapped, this is not a ``strong'' SPT phase protected by onsite symmetry $G_0$ only, and hence there exists a finite-depth quantum circuit\cite{Chen2013} $\hat U$
\begin{equation}
\hat U\dket{\psi}=\dket{T}
\end{equation}
which continuously evolves the HOSPT state $\dket{\psi}$ into a trivial product state $\dket{T}$, while preserving onsite symmetry $G_0$. We label the finite depth of circuit $\hat U$ as $d_U$.

As illustrated in FIG.\ref{fig:disentangle}, next we divide the total system $\mathcal{A}$ into two regions: its (simply-connected) interior $\mathcal{B}_0$ (both white and gray in FIG.\ref{fig:disentangle}), and boundary $\bar{\mathcal{B}_0}=\mathcal{A}\setminus\mathcal{B}_0$. We can then define a finite-depth ($d_U$) quantum circuit $U_{\mathcal{B}_0}=P_{\mathcal{B}_0}\hat U P_{\mathcal{B}_0}$ by restricting quantum circuit $\hat U$ into region $\mathcal{B}_0$, such that
\bea
\hat U_{\mathcal{B}_0}\dket{\psi}=\dket{T_{\mathcal{B}}}\otimes\dket{\psi_{\bar{\mathcal{B}}}}
\eea
where $\mathcal{B}\subset\mathcal{B}_0$ is the interior (white in FIG. \ref{fig:disentangle}) of $\mathcal{B}_0$, differing from $\mathcal{B}_0$ only by a ``cushion'' region (gray in FIG. \ref{fig:disentangle}) whose width is of the order $\sim d_U$. Here $\dket{T_{\mathcal{B}}}$ denotes the trivial product state on region $\mathcal{B}$. In other words, finite-depth quantum circuit $U_{\mathcal{B}_0}$ can continuously tune the interior region $\mathcal{B}$ of HOSPT phase into a trivial product state without closing the gap or breaking onsite symmetry $G_0$, while keeping the boundary states (on $\bar{\mathcal{B}_0}$) untouched. As a result, through finite-depth quantum circuit $\hat U_{\mathcal{B}_0}$ which preserves onsite symmetry $G_0$, the HOSPT ground state is disentangled into a trivial product state $\dket{T_{\mathcal{B}}}$ in the bulk $\mathcal{B}$, and a state $\dket{\psi_{\bar{\mathcal{B}}}}$ on its $(d-1)$-dimensional surface $\bar{\mathcal{B}}$.

Notice that in addition to preserving onsite symmetry $G_0$, the $(d-1)$-dimensional state $\dket{\psi_{\bar{\mathcal{B}}}}$ is mostly gapped except for hosting gapless modes on its $(d-k)$-dimensional submanifolds. Therefore, the gapless corner/hinge states in a HOSPT can be viewed as gapless $(d-k)$-dimensional defects on a gapped $(d-1)$-dimensional surface state $\dket{\psi_{\bar{\mathcal{B}}}}$ with onsite symmetry $G_0$. As argued in \Ref{Teo2010a,Wen2012}, the classification of such a defect falls in the classification of a $(d+1-k)$-dimensional SPT phases protected by the same onsite symmetry $G_0$.

For example, in a 2nd-order SPT phase in $d=k=2$, the gapless corner states in \eg FIG. \ref{fig:disentangle} can be viewed as gapless 0-dimensional domain walls on the gapped 1d edge. Therefore they are reduced to the 1-dimensional $G_0$-SPT phases. Similarly for a 2nd-order SPT phase in $d=3$, the gapless hinge states can be viewed as gapless 1d domain walls on a gapped 2d surface, therefore related to 2-dimensional $G_0$-SPT phases. For a 3rd-order SPT phase in $d=k=3$, the gapless corner states should be viewed as gapless 0-dimensional point defects on the gapped 2d surface with symmetry $G_0$, hence reduced to 1-dimensional $G_0$-SPT phases.

\subsection{Building HOSPT phases from lower-dimensional SPT phases}\label{sec:argument B}

In the previous argument, we have shown that the gapless $(d-k)$-dimensional boundary states in a $k$-th order SPT phase in $d$ dimensions can be reduced to the classification of $(d+1-k)$-dimensional SPT phases preserved only by onsite symmetry $G_0$. However, not all of the $G_0$-SPT phases can lead to a gapped HOSPT phase that preserves crystalline symmetry $G_c$: certain compatibility conditions must be satisfied to ensure a gapped bulk. Here we provide another argument based on the dimensional reduction approach\cite{Song2017a,Lu2017c,Huang2017b}, which explicitly builds the $k$-th order SPT phases in $d$ dimensions out of $(d-k)$-dimensional $G_0$-SPT phases.

\begin{figure}
\includegraphics[width=150pt]{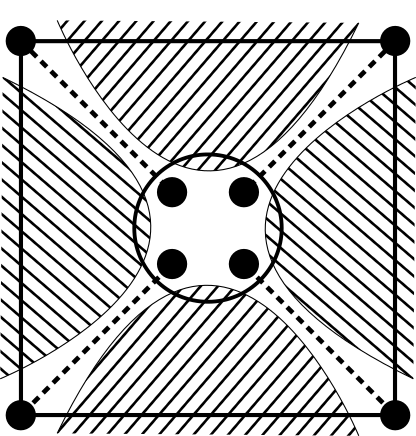}
\caption{Dimensional reduction analysis of a 2nd-order SPT phase with $G_c=C_4$ crystalline symmetry, where the black dots represent robust corner zero modes protected by onsite symmetry $G_0$. Shaded regions are trivialized by the action of a local, finite-depth quantum circuit. Dashed lines represent ``effective'' one-dimensional (1d) $G_0$-SPTs, as building blocks for the 2nd-order SPT phase in two dimensions. Note that the four endpoints of the 1d $G_0$-SPT phases must fuse to a linear representation in the bulk (circle in the middle), imposing a compatibility condition on the topological index of 1d $G_0$-SPT phases.}\label{fig:C4}
\end{figure}

Without loss of generality, we demonstrate this dimensional reduction argument using the 2nd-order 2d SPT phase with $G_c=C_4$ point group symmetry, as shown in FIG. \ref{fig:C4}). We first divide the whole open manifold $\mathcal{A}$ into 4 disconnected shaded regions $\{R_i|1\leq i\leq4\}$ in FIG. \ref{fig:C4} which are related by $C_4$ symmetry, while both the $C_4$ inversion center and 4 gapless corners lie in the rest of the space $\mathcal{A}\setminus(\bigcup_iR_i)$. Following the same construction as used in the previous argument, we can construct a $G_0$-preserving finite-depth quantum circuit $\hat U_{R_1}$ by restricting circuit $\hat U$ in region $R_1$, such that
\begin{equation}
\hat U_{R_1}\left|\psi\right\rangle = \left|T_{R_1}\right\rangle \otimes \left|\psi_{\bar{R}_1}\right\rangle
\end{equation}
where $\dket{T_{R_1}}$ represents the trivial product state on region $R_1$. By symmetrizing circuit $U_{R_1}$ w.r.t. $C_4$ rotations, we can construct a symmetric finite-depth quantum circuit
\begin{equation}
U^{sym}_R = \prod_{i=0}^3(C_4)^i\hat U_{R_1}(C_4)^{-i},~~~R\equiv\bigcup_iR_i.
\end{equation}
which preserves both onsite symmetry $G_0$ and crystalline symmetry $G_c$, such that
\begin{equation}
\hat U^{sym}_R\left|\psi\right\rangle = \left|T_R\right\rangle \bigotimes\dket{\psi_{\bar R}}
\end{equation}
In other words, symmetric finite-depth circuit $U^{sym}_R$ trivializes most of the manifold $\mathcal{A}$, except for the four 1d systems connecting the gapless corner to the rotation center. As argued previously, now that each corner state carries a projective representations of onsite symmetry $G_0$ as the boundary state of a 1d $G_0$-SPT phase, each 1d system connecting the corner to the $C_4$ rotation center must be a 1d $G_0$-SPT phases with a topological index
\bea
\nu\in\gpc{2}{G_0}{U(1)}.
\eea
Note that as a part of the gapped bulk, the $C_4$ rotation center where the ends of the four 1d $G_0$-SPT chains must form a linear representation of onsite symmetry $G_0$ \ie
\bea
4\nu\simeq 0\in\gpc{2}{G_0}{U(1)}.
\eea
This compatibility condition comes from the fusion of a number of edges of 1d SPTs dictated by the crystal symmetry $G_c=C_4$:
\begin{equation}
\label{fusion_map}
\phi:~\gpc{d+1-k}{G_0}{U(1)}\rightarrow \gpc{d+1}{G}{U(1)}
\end{equation}
Physically, the fusion map $\phi$ encodes a notion of compatibility between onsite symmetry $G_0$ and crystalline symmetry $G_c$, so that the bulk of the full system is trivial and gapped. Constructing the map $\phi$ is generally a difficult mathematical problem for an arbitrary symmetry group $G$ with both onsite and crystalline symmetries. In this paper we consider the simplest case, where the symmetry group $G = G_0 \times G_c$ is a direct product of onsite symmetry $G_0$ and global symmetry $G_c$. As we will show later, this allows a direct reduction via the K\"unneth formula, where the compatibility conditions between $(d-k)$-dimensional $G_0$-SPT phases and crystalline symmetry $G_c$ in $d$ spatial dimensions are captured by group cohomology formula (\ref{classification}).

Finally, we recall that certain SPT phases are beyond the group cohomology classification, such as the 3d time-reversal-SPT phase with $e$f$m$f surface topological orders\cite{Vishwanath2013,Wang2013a} classified by cobordism\cite{Kapustin2014,Kapustin2014a} and Kitaev's chiral 2d $E_8$ state\cite{Kitaev2006,Lu2012a}. We have also considered these beyond-group-cohomology HOSPT phases built from the $E_8$ state, as highlighted in red in TABLE \ref{tab:k=1,d=3}.

\section{Classification and construction from K\"unneth formula}

\subsection{General classification of HOSPT phases}\label{sec:classification}

In this work, we will focus on the cases where the total symmetry group $G$ is a direct product of crystalline symmetry $G_c$ and onsite symmetry $G_0$:
\bea
G=G_c\times G_0.
\eea
In this situation, there is a simple mathematical formula based on group cohomology, which gives the full classification of higher-order SPT phases.  It has been shown\cite{Jiang2017,Thorngren2018} within the group cohomology classification of SPT phases, that all $G_s$-symmetry protected topological phases of interacting bosons in $d$ spatial dimensions is given by
\bea
\gpc{d+1}{G^\ast}{U(1)}=\gpc{d+1}{G_c^\ast\times G_0}{U(1)}
\eea
where $G_c^\ast$ is isomorphic to $G_c$, obtained by replacing each orientation-reversing element of crystalline symmetry group $G_c$ by an anti-unitary operation of the same rank. According to the K\"unneth formula for group cohomology\cite{Wen2014,Wen2015,Cheng2016} we have
\bea
\notag&\gpc{d+1}{G_c^\ast\times G_0}{U(1)}=\gpc{d+2}{G_c^\ast\times G_0}{\mbz}\\
\notag&=\bigoplus_{k=0}^{d+2}\gpc{k}{G_c^\ast}{{\gpc{d+2-k}{G_s}{\mbz}}}\\
\notag&=\gpc{d+1}{G_c^\ast}{U(1)}\oplus\gpc{d+1}{G_c^\ast}{\gpc{1}{G_0}{\mbz}}\\
&\bigoplus_{k=0}^{d}\gpc{k}{G_c^\ast}{{\gpc{d-k+1}{G_0}{U(1)}}}.\label{kunneth formula}
\eea
The 1st term $\gpc{d+1}{G_c^\ast}{U(1)}$ classifies crystalline SPT phases protected only by crystalline symmetry $G_c$\cite{Song2017a,Thorngren2018}. The 2nd term $\gpc{d+1}{G_c^\ast}{\gpc{1}{G_0}{\mbz}}$ vanishes for any finite group\cite{Cheng2016}, as in the case considered here where $G_c$ is a point group or magnetic point group.

Therefore we shall focus on the last line of the above K\"unneth formula (\ref{kunneth formula}). Each term in
\bea\label{class:(k+1)-order in d-dim}
\gpc{k}{G_c^\ast}{{\gpc{d-k+1}{G_0}{U(1)}}},~~~0\leq k\leq d.
\eea
can be interpreted as the classification of $(k+1)$-th order SPT phases in $d$ spatial dimensions, protected by onsite symmetry $G_0$ and cystalline symmetry $G_c$. Such a SPT phase is featured by robust gapless states on proper $(d-k-1)$-dimensional open boundaries, which are protected by onsite symmetry $G_0$ alone. For example, the $k=0$ term in (\ref{class:(k+1)-order in d-dim})
\bea
\gpc{0}{G_c^\ast}{{\gpc{d+1}{G_0}{U(1)}}}=\gpc{d+1}{G_0}{U(1)}
\eea
corresponds to the 1st-order (\ie the usual ``strong'') SPT phases protected by onsite symmetry $G_0$, featured by gapless modes on $(d-1)$-dimensional boundaries.

2nd-order SPT phases in $d\geq2$ are all captured by $k=1$ term in (\ref{class:(k+1)-order in d-dim})
\bea\label{class:2nd order}
\gpc{1}{G_c^\ast}{{\gpc{d}{G_0}{U(1)}}}
\eea
which host gapless (or anomalous topological orders when $d\geq4$) excitations on $(d-2)$-dimensional boundaries protected by onsite symmetry $G_0$.

Similarly, 3rd-order SPT phases in $d\geq3$ are all captured by $k=2$ term in (\ref{class:(k+1)-order in d-dim})
\bea\label{class:3rd order}
\gpc{2}{G_c^\ast}{{\gpc{d-1}{G_0}{U(1)}}}
\eea
which host gapless (or anomalous topological orders when $d\geq5$) excitations on $(d-3)$-dimensional boundaries protected by onsite symmetry $G_0$, such as corner states in $d=3$.

\subsection{``Strong'' HOSPT phases versus ``weak'' crystalline SPT phases}

As mentioned previously, we define $k$-th order SPT phases in $d$ dimensions by the presence of robust $(d-k)$-dimensional topological boundary states, protected by onsite (or global) symmetry $G_0$ only. These are ``strong'' SPT phases, whose boundary excitations do not require protection from the crystalline symmetry $G_c$. In comparison, there are also ``weak'' crystalline SPT phases, whose topological boundary excitations are protected by crystalline symmetries (in addition to onsite symmetries)\cite{Fu2007,Ran2010,Song2015,Thorngren2018}. In fact, in addition to strong HOSPT phases which are the focus of this paper, certain weak crystalline SPT phases are encoded inside the whole Kunneth formula (\ref{kunneth formula}), such as those colored in green in TABLE \ref{tab:k=2,d=3}. Before systematically analyzing and constructing HOSPT phases in detail, we briefly discuss the weak crystalline SPT phases.

First of all, the $k=d+1$ term $\gpc{d+1}{G_c^\ast}{U(1)}$ in Kunneth formula clearly describes weak SPT phases protected only by the crystalline symmetry $G_c$.

Next, we comment on $k=d$ term in (\ref{class:(k+1)-order in d-dim}):
\bea\label{class:(d+1)-order}
\gpc{d}{G_c^\ast}{\gpc{1}{G_0}{U(1)}}.
\eea
The physics of this term is to assign onsite symmetry charges (linear representation $\gpc{1}{G_0}{U(1)}$ of onsite symmetry $G_0$) to defects of the crystalline symmetry $G_c$. In a simplest example, for the $k=d=1$ case of 1d insulators ($G_0=U(1)$) with inversion symmetry $I$ ($G_c=Z_2^{\bsi}$), we have $\gpc{1}{U(1)}{U(1)}=\mbz$ and hence
\bea
\gpc{1}{G_c^\ast}{\gpc{1}{U(1)}{U(1)}}=\gpc{1}{Z_2^\bst}{\mbz}=\mbz_2
\eea
The nontrivial element of the above $\mbz_2$ classification corresponds to assigning an odd number of $U(1)$ charges to the inversion center, while the trivial element corresponds to having an even number of $U(1)$ charge on the inversion center. There is no gapless boundary excitations for either of the two phases in 1d.

However in $k=d\geq2$, weak SPT phases with boundary states protected by crystalline symmetry generally can appear in the Kunneth formula (\ref{kunneth formula}). For example in $k=d=2$ case with mirror symmetry $G_c=Z_2^\bsm$, $\gpc{2}{Z_2^\bsm}{\gpc{1}{G_0}{U(1)}}$ corresponds to assigning $G_0$ charges to each domain wall of mirror symmetry $\bsm$ on the 1d mirror axis of the 2d system. This leads to gapless boundary states if the boundary of the system preserves mirror symmetry. Similarly in $d=k=3$ case with $n$-fold rotational symmetry $G_c=C_n$, $\gpc{3}{C_n}{\gpc{1}{G_0}{U(1)}}$ corresponds to assigning $G_0$ charges to each domain wall of $C_n$ rotational symmetry on the 1d rotation axis. This leads to weak 3d crystalline SPT phases, hosting gapless (or anomalous) boundary states if the boundary preserves $C_n$ symmetry.

Another example is $k=d-1$ in (\ref{class:(k+1)-order in d-dim}). Take $d=3,~k=2$ for instance, considering mirror symmetry $G_c=Z_2^\bsm$ again, $\gpc{2}{Z_2^\bsm}{\gpc{2}{G_0}{U(1)}}$ corresponds to assigning 1d $G_0$-SPT phases classified by ${\gpc{2}{G_0}{U(1)}}$ to each mirror domain wall on the 2d mirror plane. This can lead to gapless (or anomalous) boundary states protected by both mirror and onsite $G_0$ symmetry, if the boundary preserves mirror symmetry $\bsm$.

As we mentioned before, the boundary states of these weak crystalline SPT phases will generally be destroyed by perturbations that break the crystalline symmetry, such as disorders and crystalline distortions. Meanwhile, their interpretation in the Kunneth formula can be quite tricky, as shown in the above examples. Hereafter we will be focusing on the strong HOSPT phases, whose topological boundary excitations are robust even if crystalline symmetries are broken on the surface.

\subsection{Decorated domain wall construction}\label{sec:ddw}

Here we briefly describe how to explicitly construct the higher-order SPT phases in $d$ spatial dimensions, using $G_0$-SPT phases in lower dimensions. In particular, the group cohomology formula (\ref{class:(k+1)-order in d-dim}) provides a clear physical meaning for such a construction, similar to the decorated domain wall construction\cite{Chen2014} for the usual (``1st-order'') SPT phases.

First we consider 2nd-order SPT phases in $d$ dimensions, classified by 1st group cohomology
\bea\notag
&\{\nu_1(g_0,g_1)\in\gpc{d}{G_0}{U(1)}|g_i\in G_c^\ast\}\\
&\in\gpc{1}{G_c^\ast}{\gpc{d}{G_0}{U(1)}}
\eea
These are nothing but linear representations of the symmetry group $G_c^\ast$
\bea
U_g\equiv\nu_1(1,g)\in\gpc{d}{G_0}{U(1)},~~~g\in G_c^\ast.
\eea
satisfying the 1-cocycle condition
\bea
&U_g\cdot U^{s(g)}_h=U_{g\cdot h},\\
&\notag s(g)=\pm1~\text{for}~g=\text{unitary/anti-unitary}.
\eea
$U_g$ valued in $\gpc{d}{G_0}{U(1)}$ physically represents a domain wall labeled by symmetry element $g$, decorated by $(d-1)$-dimensional $G_0$-SPT phases labeled by elements in $\gpc{d}{G_0}{U(1)}$. The above 1-cocycle condition can be viewed as a compatibility condition between the addition rules of $(d-1)$-dimensional $G_0$-SPT phases and the addition rules ($g\cdot h=gh$) of domain walls, in order to ensure a gapped bulk spectrum. To understand this, we see that a domain wall of the $G_{c}^{\ast}$ symmetry is labeled by a group element $g_1 \in G_c^\ast$.  The $(d-1)$-dimensional SPT phase associated with this domain wall is labeled by an element $m_1 \in\gpc{d}{G_0}{U(1)}$.  The fusion of two domain walls $g_1\cdot g_2$ combines these $(d-1)$-dim $G_0$-SPT's into $m_1 + m_2$. However, the fusion must respect the group structure of $\gpc{d}{G_0}{U(1)}$, and this consistency condition is exactly captured by Eq. (\ref{class:(k+1)-order in d-dim}). Therefore each element of $\gpc{1}{G_c^\ast}{\gpc{d}{G_0}{U(1)}}$ describes a way to assign $(d-1)$-dimensional $G_0$-SPT phases on the domain walls of crystalline symmetry $G_c$, which is compatible with a gapped bulk.

Next we consider 2nd-order SPT phases in $d$ dimensions, classified by 2nd group cohomology
\bea\notag
&\{\nu_2(g_0,g_1,g_2)\in\gpc{d-1}{G_0}{U(1)}|g_i\in G_c^\ast\}\\
&\in\gpc{2}{G_c^\ast}{\gpc{d-1}{G_0}{U(1)}}
\eea
They are nothing but projective representation of symmetry group $G_c^\ast$
\bea
&U_g\cdot U_h^{s(g)}=\omega(g,h) U_{gh},~~~g,h\in G_c^\ast;\\
&\omega(g,h)\equiv\nu_2(1,g,gh)\in\gpc{d-1}{G_0}{U(1)}.
\eea
satisfying the 2-cocycle (or associativity) condition
\bea
\omega(g,h)\omega(gh,k)=\omega(g,hk)\omega^{s(g)}(h,k),~~~g,h,k\in G_c^\ast.
\eea
Since $U_g$ represents the $(d-1)$-dimensional domain wall labeled by element $g$ of crystalline symmetry $G_c^\ast$, $\omega(g,h)$ naturally represents the $(d-2)$-dimensional manifold where three domain walls $U_g$, $U_h$ and $U_{(gh)^{-1}}$ intersect. The fact that $\omega(g,h)$ takes values in $\gpc{d-1}{G_0}{U(1)}$ physically means that these domain wall intersections are decorated by $(d-2)$-dimensional $G_0$-SPT phases, which are classified by group cohomology $\gpc{d-1}{G_0}{U(1)}$.

As a simplest example, we consider the $n$-fold rotational symmetry $G_c=C_n$. Each of the $n$ domain walls of the $C_n$ symmetry can be decorated by the same $(d-1)$-dimensional $G_0$-SPT phase, such that $n$ copies of these $G_0$-SPT phases intersect at the $C_n$ rotational axis. For the system to be gapped on the rotational axis, these $n$ copies of $G_0$-SPT phases together must fuse to a trivial phase with no gapless boundary states. This exactly corresponds to 2nd-order SPT phases classified by $\gpc{1}{C_n^\ast\simeq Z_n}{\gpc{d}{G_0}{U(1)}}$. Meanwhile at the intersection of $n$ domain walls of $C_n$ symmetry, the rotational axis itself can also be decorated by a $(d-2)$-dimensional $G_0$-SPT phases, which corresponds to the 3rd-order SPT phases classified by $\gpc{2}{C_n^\ast\simeq Z_n}{\gpc{d-1}{G_0}{U(1)}}$.

Another example is the mirror reflection symmetry $G_c=Z_2^\bsm$, where the orientation-reversing mirror symmetry $\bsm$ should be regarded as an anti-unitary symmetry when computing the group cohomology. For $k=d=2$, the 2nd-order SPT phases in classified by $\gpc{1}{Z_2^\bsm}{\gpc{2}{G_0}{U(1)}}$ can be understood as assigning a 1d $G_0$-SPT phases on each mirror plane.

Below we will classify 2nd-order SPT phases in $d=2,3$ (TABLE \ref{tab:k=1,d=2},\ref{tab:k=1,d=3}) and 3rd-order SPT phases in $d=3$ (TABLE \ref{tab:k=2,d=3}), for various choices of onsite symmetry $G_0$ and crystalline (and magnetic crystalline) symmetry $G_c$. We will also explicitly construct these higher-order SPT phases using the decorated domain wall picture as described above, in section \ref{sec:k=1,d=2}-\ref{sec:k=2,d=3}.

\begin{table*}[tb]
\centering
\begin{tabular} {|c||c|c|c|c|c|}
\hline
$k=1,d=2$&\multicolumn{5}{c|}{Onsite symmetry $G_0$}\\
\hline
&$Z_2^\bst$&$SO(3)$ or &$SO(3)\times Z_2^\bst$ &$Z_a\times Z_b$&$Z_a\times Z_2^\bst$\\
Crystalline symmetry $G_c$&&$U(1)\rtimes Z_2$&or $U(1)\times Z_2^\bst$&&\\
\hline
$C_n$&$\mbz_{(n,2)}$&$\mbz_{(n,2)}$&$\mbz_{(n,2)}$&$\mbz_{(n,a,b)}$&$\mbz_{(n,2)}\times\mbz_{(n,a,2)}$\\
\hline
$C_{n,\text{v}}=C_n\rtimes Z_2^{\bsm_\text{v}}$&$\mbz_{(n,2)}\times\mbz_2$&$\mbz_{(n,2)}\times\mbz_2$&$\mbz_{(n,2)}^2\times\mbz_2^2$&$\mbz_{(n,a,b)}\times\mbz_{(2,a,b)}$&$\mbz_{(n,2)}\times\mbz_{(n,a,2)}\times\mbz_{2}\times\mbz_{(a,2)}$\\
\hline
$C_{2n}^\bst\equiv\{(c_{2n}\cdot\bst)^m|0\leq m<2n\}$&$\mbz_{2}$&$\mbz_{2}$&$\mbz_{2}^2$&$\mbz_{(2,a,b)}$&$\mbz_{2}\times\mbz_{(a,2)}$\\
\hline
$C_{2n}^\bst\rtimes Z_2^{\bsm_\text{v}}$&$\mbz_{2}^2$&$\mbz_{2}^2$&$\mbz_{2}^4$&$\mbz_{(2,a,b)}^2$&$\mbz_{2}^2\times\mbz_{(a,2)}^2$\\
\hline
$C_{n}\rtimes Z_2^{\bsm_\text{v}\cdot\bst}$&$\mbz_{(n,2)}\times\mbz_{2}$&$\mbz_{(n,2)}\times\mbz_{2}$&$\mbz_{(n,2)}^2\times\mbz_2^2$&$\mbz_{(2,n,a,b)}\times\mbz_{(2,a,b)}$&$\mbz_{(n,2)}\times\mbz_{(n,a,2)}\times\mbz_{2}\times\mbz_{(a,2)}$\\
\hline
\hline
\color{blue}{$d=1$ $G_0$-SPTs: $\gpc{2}{G_0}{U(1)}$}&\color{blue}{$\mbz_2$}&\color{blue}{$\mbz_2$}&\color{blue}{$\mbz_2^2$}&\color{blue}{$\mbz_{(a,b)}$}&\color{blue}{$\mbz_2\times\mbz_{(a,2)}$}\\
\hline
\end{tabular}
\caption{2nd-order bosonic SPT phases ($k=1$) in $d=2$ spatial dimensions, protected by symmetry group $G_s=G_c\times G_0$ where $G_c$ and $G_0$ represent the crystalline and onsite symmetry group respectively. The general classification is given by linear representation $\gpc{1}{G_c^\ast}{\gpc{2}{G_0}{U(1)}}$ as shown in (\ref{class:2nd order}) with $d=2$. Through a dimensional reduction procedure they are all constructed from $G_0$-SPT phases in $d=1$ dimension, classified by $\gpc{2}{G_0}{U(1)}$ in the last line (blue).}
\label{tab:k=1,d=2}
\end{table*}

\section{2nd-order SPT phases in two dimensions}\label{sec:k=1,d=2}

As shown in (\ref{class:2nd order}), the 2nd-order SPT phases in $d$ spatial dimension are classified by $\gpc{k=1}{G_c^\ast}{\gpc{d}{G_0}{U(1)}}$, \ie the linear representation of group $G_c^\ast$ whose coefficients take value in the $(d-1)$-dimensional SPT classification $\gpc{d}{G_0}{U(1)}$. For $d=2$ case, the building blocks of 2nd-order SPT phases in two dimensions are 1d SPT phases protected by onsite symmetry $G_0$. Below we provide a full classification for 2nd-order SPT phases in $d=2$ with all possible 2d point group and magnetic point group symmetries, and describe how to use 1d SPT phases to construct these 2nd-order 2d SPT phases.

\subsection{Classification}

To compute $\gpc{1}{G_c^\ast}{\gpc{d}{G_0}{U(1)}}$, first we need to obtain the group $G_c^\ast$ from crystalline symmetry $G_c$. As mentioned earlier, $G_c^\ast$ is isomorphic to $G_c$, obtained by replacing each orientation-reversing element $g$ of $G_c$ by an anti-unitary operation $g^\ast$ of the same rank. For example we have
\bea
&G_c=C_n\Longrightarrow G_c^\ast\simeq Z_n;\\
&G_c=C_n\rtimes Z_2^{\bsm_\text{v}}\Longrightarrow G_c^\ast\simeq Z_n\rtimes Z_2^\bst;\\
&G_c=C^\bst_{2n}~\text{or}~S_{2n}\Longrightarrow G_c^\ast\simeq .
\eea
where we use $Z_{2n}^\bst$ to denote a group generated by an anti-unitary operator $\bst$ of ranking $2n$.

After obtaining $G_c^\ast$, the next step is to compute $\gpc{d}{G_0}{U(1)}$, the coefficient of the desired linear representation $\mathcal{H}^1$. For $d=2$ case, $\gpc{2}{G_0}{U(1)}$ corresponds to the classification of 1d SPT phases\cite{Turner2011,Fidkowski2011,Chen2011,Schuch2011} protected by onsite symmetry $G_0$: it always forms a finite Abelian group, as summarized in the last line of TABLE \ref{tab:k=1,d=2}.

Generally the classification of SPT phases with onsite symmetry $G_0$ always form a discrete Abelian group, which holds for the group cohomology classification $\gpc{d}{G_0}{U(1)}$ and beyond. One important relation for group cohomology is
\bea\label{factorization}
\gpc{k}{G}{A\times B}=\gpc{k}{G}{A}\times\gpc{k}{G}{B}.
\eea
Therefore to compute $\gpc{1}{G_c^\ast}{\gpc{d}{G_0}{U(1)}}$ in (\ref{class:2nd order}), we only need to know $\gpc{1}{G}{\mbz}$, and $\gpc{1}{G}{\mbz_a}$ for any finite integer $a\in\mbz$. Since $\gpc{2}{G_0}{U(1)}$ is always a finite Abelian group, making use of relation (\ref{factorization}), we can compute $\gpc{1}{G_c^\ast}{\gpc{2}{G_0}{U(1)}}$ purely based on knowledge of $\gpc{1}{G}{\mbz_a}$ for any finite integer $a$. Below we list $\gpc{1}{G_c^\ast}{\mbz_a}$ for all $d=2$ point groups and magnetic point groups $G_c$:
\bea
&\notag (C_n)^\ast\simeq Z_n,~~(S_{2n}^\bst)^\ast\simeq Z_{2n},\\
\label{k=1,Za:1st}&\gpc{1}{Z_n}{\mbz_a}=\mbz_{(n,a)};\\
&\notag (C_{n,\text{v}})^\ast\simeq(C_n\rtimes Z_2^{\bsm_\text{v}\bsm_\text{h}\bst})^\ast\simeq Z_n\rtimes Z_2^\bst,\\
&\gpc{1}{Z_n\rtimes Z_2^\bst}{\mbz_a}=\mbz_{(n,a)}\times\mbz_{(2,a)};\\
&\notag(C_{2n}^\bst)^\ast\simeq(S_{2n})^\ast\simeq Z_{2n}^\bst;\\
&\gpc{1}{Z_{2n}^\bst}{\mbz_a}=\mbz_{(2,a)};\\
&\notag(D_{n,\text{d}})^\ast\simeq(C_{2n}^\bst\rtimes Z_2^{\bsm_\text{v}})^\ast\simeq Z_{2n}^\bst\rtimes Z_2,\\
&\gpc{1}{Z_{2n}^\bst\rtimes Z_2}{\mbz_a}=\mbz_{(2,a)}^2;\\
&\notag(D_n)^\ast\simeq(C_n\rtimes Z_2^{\bsm_\text{v}\bst})^\ast\simeq Z_{n}\rtimes Z_2,\\
&\gpc{1}{Z_{n}\rtimes Z_2}{\mbz_a}=\mbz_{(n,a,2)}\times\mbz_{(a,2)};\\
&\notag (C_{n,\text{h}})^\ast\simeq(C_n\times Z_2^\bsi)^\ast\simeq Z_n\times Z_2^\bst,\\
&\gpc{1}{Z_n\times Z_2^\bst}{\mbz_a}=\mbz_{(n,a,2)}\times\mbz_{(2,a)};\\
&\notag (D_{n,\text{h}})^\ast\simeq(Z_n\rtimes Z_2)\times Z_2^\bst,\\
&\gpc{1}{(Z_n\rtimes Z_2)\times Z_2^\bst}{\mbz_a}=\mbz_{(n,a,2)}\times\mbz_{(2,a)}^2.\label{k=1,Za:last}
\eea

Using relation (\ref{factorization}) and the above results (\ref{k=1,Za:1st})-(\ref{k=1,Za:last}), we acquire the classification of all 2nd-order SPT phases in $d=2$, as summarized in TABLE \ref{tab:k=1,d=2}.

\subsection{Examples}

\subsubsection{$G_c=C_n$}

\begin{figure}
\includegraphics[width=150pt]{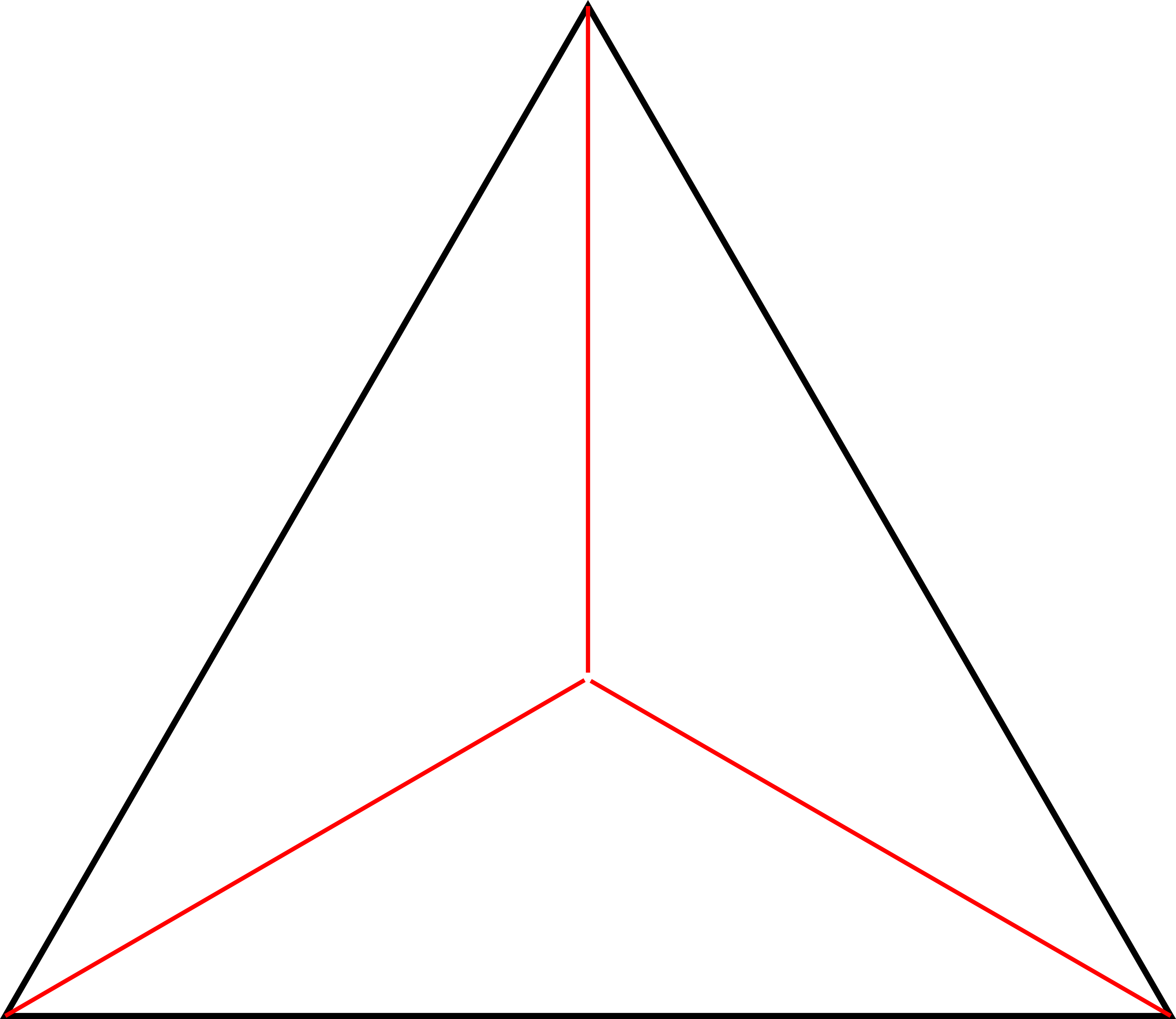}
\caption{2nd-order SPT phases with $G_c=C_3$ point group symmetry, where each of the three $C_3$ domain walls is decorated by the same 1d $G_0$-SPT phase meeting at the rotation center. The projective representations at the end of each 1d $G_0$-SPT phase must form a linear representation to ensure a gapped bulk, as manifested in group cohomology formula (\ref{class:2nd order}).}\label{fig:c3}
\end{figure}

The simplest examples of 2nd-order SPT phases are protected by $n$-fold rotational symmetry $G_c=C_n$, classified by
\bea
\gpc{1}{Z_n}{\gpc{2}{G_0}{U(1)}}
\eea
They can all be built from 1d SPT phases protected by onsite symmetry $G_0$, where the 1d $G_0$-SPT phases are aligned in a $C_n$-symmetric manner as shown in FIG. \ref{fig:c3} for $G_c=C_3$ case. Since the endpoints of $n$ copies of 1d $G_0$-SPT phases intersect at the center of the system (see FIG. \ref{fig:c3}), they must form a linear representation of onsite symmetry $G_0$ to ensure a gapped symmetric bulk. This provides a compatibility condition for the 1d $G_0$-SPT phases, manifested in the group cohomology formula
\bea\label{gpc:1,Zn,Za}
\gpc{1}{Z_n}{\mbz_a}=\mbz_{(n,a)}
\eea
where $(n,a)$ is the greatest common divisor of integers $n$ and $a$.

Formula (\ref{gpc:1,Zn,Za}) can be understood as follows. The group cohomology $\gpc{1}{Z_n}{\mbz_a}$ stands for linear representation $\{U_g|g\in Z_n\}$ of $Z_n$ group with coefficients in $\mbz_a$-valued phase factors
\bea
\mbz_a=\{u_j\equiv e^{2\pi\imth\frac{j}{a}}|0\leq \nu<a,~j\in\mbz\}
\eea
Denoting the generator of $Z_n$ group as $R$ with $R^n=1$, we have
\bea
U_R\dket{\nu}=u_{\nu}\dket{\nu}\Longrightarrow U_{1=R^n}=(U_R)^n=u_{n\nu}=1
\eea
and as a result
\bea\label{rep:Cn}
n\nu=0\mod a\Longrightarrow\nu=m\frac{a}{(n,a)},~~~0\leq m<(n,a).
\eea
where $(n,a)$ denotes the greatest common divisor of integers $a$ and $n$. Physically this means the topological index $\nu\in\mbz_a$ of the 1d $G_0$-SPT phase decorated on each $C_n$ domain wall must be a multiple of $\frac{a}{(n,a)}$, to ensure the bulk to be gapped at the rotation axis where the $n$ domain walls intersect. Hence there are $(n,a)$ distinct 2nd-order SPT phases with $C_n$ symmetry, characterized by the 1d $G_0$-SPT phase with $\nu=0,~\frac{a}{(n,a)},~2\frac{a}{(n,a)},\cdots$ on each domain wall. This corresponds to the $\mbz_{(n,a)}$ classification in formula (\ref{gpc:1,Zn,Za}).

One immediate physical consequence is the presence of zero-energy corner modes located on each corner of the $C_n$-symmetric finite system shown in FIG. \ref{fig:c3}. Each corner mode is nothing but the boundary states of 1d $G_0$-SPT phases with $\nu=0\mod\frac{a}{(n,a)}$, which carries a projective representation of onsite symmetry $G_0$. Notice that with only $C_n$ symmetry, the 1d $G_0$-SPT phases can together be rotated around the $C_n$ center by an arbitrary angle, and therefore the zero-energy corner states will only appear in certain (but not all) finite systems.

\subsubsection{$G_c=C_{n,\text{v}}\equiv C_n\rtimes Z_2^{\bsm_\text{v}}$ or $C_n\rtimes Z_2^{\bsm_\text{v}\bsm_\text{h}\bst}$}

\begin{figure}
\includegraphics[width=150pt]{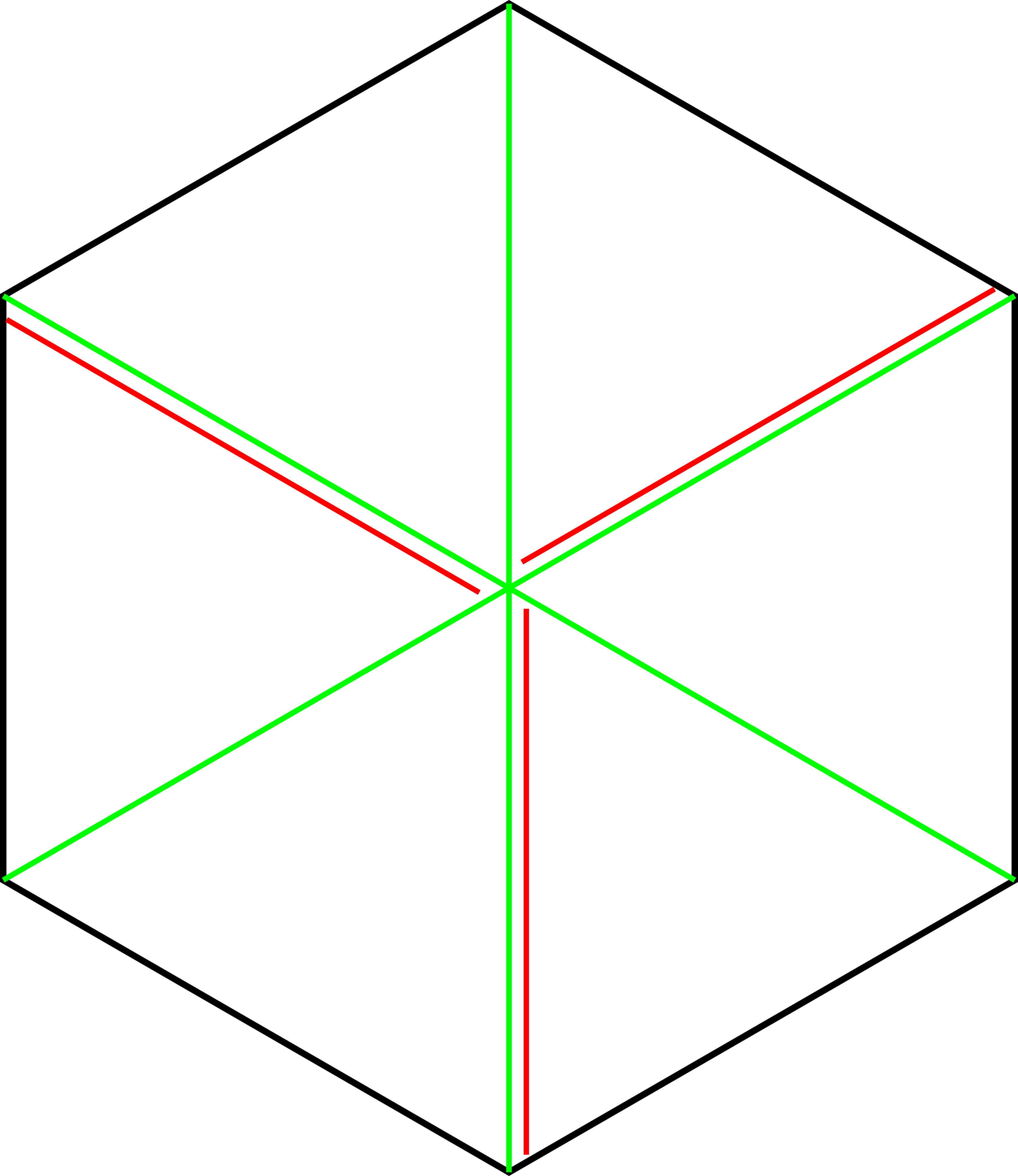}
\caption{2nd-order SPT phases with $G_c=C_{3,\text{v}}$ point group symmetry, where each of the three $C_3$ domain walls is decorated by the same 1d $G_0$-SPT phase meeting at the rotation center, illustrated by red lines. Meanwhile each mirror plane can also be decorated by another 1d $G_0$-SPT phase, labeled by the green lines.}\label{fig:c3m}
\end{figure}

Consider point group $G_c=C_{n,\text{v}}$, generated by $n$-fold rotation $R$ with $R^n=1$ along $\hat z$-axis, and mirror reflection $\bsm_\text{v}$ whose mirror plane is parallel to $\hat z$-axis. As described earlier, the associated 2nd-order SPT phases are classified by the linear representation (1st group cohomology) of $(C_{n,\text{v}})^\ast\simeq Z_n\rtimes Z_2^\bst$, with coefficients in 1d $G_0$-SPT phases classified by $\gpc{2}{G_0}{U(1)}$. The decorated-domain-wall construction of these 2nd-order SPT phases with $C_{n,\text{v}}$ symmetry can be implied from the following formula
\bea\label{gpc:1,Cnv,Za}
\gpc{1}{Z_n\rtimes Z_2^\bst}{\mbz_a}=\mbz_{(n,a)}\times\mbz_{(2,a)}
\eea
The 1st factor $\mbz_{(n,a)}$ labels the 1d $G_0$-SPT phases assigned on each $C_n$ domain walls and intersected at the $C_n$ rotation center, illustrated by the red lines in FIG. \ref{fig:c3m}. On the other hand, the 2nd factor $\mbz_{(2,a)}$ labels the 1d $G_0$-SPT phases placed on each of the $n$ mirror planes, illustrated by the green lines in FIG. \ref{fig:c3m}. The linear representation $\{U_R,U_{\bst}\}$ corresponding to $\gpc{1}{Z_n\rtimes Z_2^\bst}{\mbz_a}$ satisfies the following algebraic conditions:
\bea
(U_R)^n=U_\bst U_\bst^\ast=U_RU_\bst(U_RU_\bst)^\ast=1.
\eea
Similar to $G_c=C_n$ case discussed earlier, in (\ref{gpc:1,Cnv,Za}) the linear representation of $n$-fold rotation $U_R$ is given by
\bea
U_R\dket{\nu}=u_{\nu_R}\dket{\nu},~~u_{\nu_R}=e^{2\pi\imth\frac{\nu_R}{a}},~~\nu_R=0\mod\frac{a}{(n,a)}.
\eea
While $U_R$ is invariant under any gauge transformation on the basis vectors of the linear representation, this is not the case for anti-unitary operator $\bst=\bsm_\text{v}^\ast$. Specifically under a gauge rotation by phase factor $e^{\frac{2\pi\imth}{a}}$ on all basis vectors, the linear representation of anti-unitary operator $\bst$ changes as
\bea
\notag&U_\bst\dket{\nu}=u_{\nu_\bst}\dket{\nu},~~\dket{\nu}\rightarrow e^{\frac{2\pi\imth}{a}}\dket{\nu}\Longrightarrow\\
&U_\bst=u_{\nu_\bst}\rightarrow e^{\frac{2\pi\imth}{a}}u_{\nu_\bst}(e^{-\frac{2\pi\imth}{a}})^\ast=u_{\nu_\bst+2}.
\eea
This indicates that 1d $G_0$-SPT index $\nu_\bst$ on each mirror plane is only well-defined modulo 2, leading to the $\mbz_{(2,a)}$ factor in formula (\ref{gpc:1,Cnv,Za}). This result has a straightforward physical interpretation: two 1d $G_0$-SPT phases of the same topological index can be merged from two sides into the mirror plane, hence changing the 1d topological index on the mirror plane by any even integer without closing the bulk gap.

Unlike in the previous $G_c=C_n$ case where the $n$ copies of 1d $G_0$-SPT phases can be rotated by an arbitrary angle, here due to the presence of $n$ mirror planes (related by $C_n$ rotations), all 1d $G_0$-SPT phases are assigned to the mirror planes. As a result, as long as the corners of the finite system lie on the mirror planes, they will give rise to zero-energy corner modes protected by onsite $G_0$ symmetry. However as illustrated in FIG. \ref{fig:c3m}, there are two different types of corners, terminating the green lines only versus terminating both green lines. These two types of corners generally support different types of projective representations of onsite symmetry $G_0$.

Finally, it is straightforward to show that the above classification and construction remain true for magnetic point group $G_c=C_n\rtimes Z_2^{\bsm_\text{v}\bsm_\text{h}\bst}$, generated by rotation $C_n$ around $\hat z$-axis and 2-fold anti-unitary magnetic rotation $\bsm_\text{v}\bsm_\text{h}\bst$ around an in-plane (such as $\hat x$) axis.

\begin{table*}[tb]
\centering
\begin{tabular} {|c||c|c|c|c|c|}
\hline
$k=1,d=3$&\multicolumn{5}{c|}{Onsite symmetry $G_0$}\\
\hline
&$U(1)$&$U(1)\rtimes Z_2$&$SO(3)\times Z_2^\bst$ &$Z_a$&$Z_a\times Z_2^\bst$\\
Crystalline symmetry $G_c$&or $SO(3)$&&&&\\
\hline
$C_n$&$\mbz_1$&$\mbz_{(n,2)}$&$\mbz_{(n,2)}$&$\mbz_{(n,a)}$&$\mbz_{(n,a,2)}^2$\\
\hline
$C_{n,\text{v}}$ or $C_n\rtimes Z_2^{\bsm_\text{v}\bsm_\text{h}\bst}$&$\mbz_2\times\alert{\mbz_2}$&$\mbz_2^2\times\mbz_{(n,2)}\times\alert{\mbz_2}$&$\mbz_{(n,2)}\times\mbz_2$&$\mbz_{(n,a)}\times\mbz_{(2,a)}\times\alert{\mbz_2}$&$\mbz_{(n,a,2)}^2\times\mbz_{(a,2)}^2$\\
\hline
$C_{n,\text{h}}\equiv C_n\times Z_2^{\bsm_\text{h}}$&$\mbz_2\times\alert{\mbz_2}$&$\mbz_2^2\times\mbz_{(n,2)}\times\alert{\mbz_2}$&$\mbz_{(n,2)}\times\mbz_2$&$\mbz_{(n,a,2)}\times\mbz_{(2,a)}\times\alert{\mbz_2}$&$\mbz_{(n,a,2)}^2\times\mbz_{(a,2)}^2$\\
\hline
$D_{n}\equiv C_n\rtimes Z_2^{\bsm_\text{h}\cdot\bsm_\text{v}}$ or $C_{n}\rtimes Z_2^{\bsm_\text{v}\cdot\bst}$&$\mbz_1$&$\mbz_{(n,2)}\times\mbz_2$&$\mbz_{(n,2)}\times\mbz_2$&$\mbz_{(n,a,2)}\times\mbz_{(2,a)}$&$\mbz_{(n,a,2)}^2\times\mbz_{(a,2)}^2$\\
\hline
$D_{n,\text{h}}\equiv C_{n,\text{v}}\times Z_2^{\bsm_\text{h}}$&$\mbz_2\times\alert{\mbz_2}$&$\mbz_2^3\times\mbz_{(n,2)}\times\alert{\mbz_2}$&$\mbz_{(n,2)}\times\mbz_2^2$&$\mbz_{(n,a,2)}\times\mbz_{(2,a)}^2\times\alert{\mbz_2}$&$\mbz_{(n,a,2)}^2\times\mbz_{(a,2)}^4$\\
\hline
$C_{2n}^\bst$ or $S_{2n}\equiv\{(c_{2n}\cdot\bsm_\text{h})^m|0\leq m<2n\}$&$\mbz_2\times\alert{\mbz_2}$&$\mbz_2^2\times\alert{\mbz_2}$&$\mbz_2$&$\mbz_{(2,a)}\times\alert{\mbz_2}$&$\mbz_{(2,a)}^2$\\
\hline
$D_{n,\text{d}}\equiv S_{2n}\rtimes Z_2^{\bsm_\text{v}}$ or $C_{2n}^\bst\rtimes Z_2^{\bsm_\text{v}}$&$\mbz_2\times\alert{\mbz_2}$&$\mbz_2^3\times\alert{\mbz_2}$&$\mbz_2^2$&$\mbz_{(2,a)}^2\times\alert{\mbz_2}$&$\mbz_{(2,a)}^4$\\
\hline
$S_{2n}^\bst\equiv\{(c_{2n}\cdot\bsm_\text{h}\cdot\bst)^m|0\leq m<2n\}$&$\mbz_1$&$\mbz_2$&$\mbz_2$&$\mbz_{(2n,a)}$&$\mbz^2_{(a,2)}$\\
\hline
%$C_{2n}^\bst\rtimes Z_2^{\bsm_\text{v}}$&&&&&\\
%\hline
%$C_{n}\rtimes Z_2^{\bsm_\text{v}\cdot\bst}$&&&&&\\
%\hline
\hline
$d=2$ $G_0$-SPTs: $\gpc{3}{G_0}{U(1)}$ plus \color{red}{$E_8$ state}&$\mbz\times\alert{\mbz}$&$\mbz\times\mbz_2\times\alert{\mbz}$&$\mbz_2$&$\mbz_{a}\times\alert{\mbz}$&$\mbz^2_{(a,2)}$\\
\hline
\end{tabular}
\caption{(color online) 2nd-order bosonic SPT phases ($k=1$) in $d=3$ spatial dimensions, protected by symmetry group $G_s=G_c\times G_0$ where $G_c$ and $G_0$ represent the crystalline and onsite symmetry group respectively. The general classification is given by linear representation $\gpc{1}{G_c^\ast}{\gpc{3}{G_0}{U(1)}}$ as shown in (\ref{class:2nd order}) with $d=3$, except for the ``beyond-cohomology'' states colored by red. Through a dimensional reduction procedure they can all be built from $G_0$-SPT phases in $d=2$ dimension, classified by $\gpc{3}{G_0}{U(1)}$ in the last line. Red-colored ``beyond-cohomology'' states are built from the chiral bosonic $E_8$ state\cite{Kitaev2006,Lu2012a}.}
\label{tab:k=1,d=3}
\end{table*}

\section{2nd-order SPT phases in three dimensions}\label{sec:k=1,d=3}

\subsection{Classification}

2nd-order SPT phases in $d=3$ are classified by $\gpc{1}{G_c^\ast}{\gpc{3}{G_0}{U(1)}}$, \ie linear representation of group $G_c^\ast$ with coefficients valued in $\gpc{3}{G_0}{U(1)}$. Physically this means the building blocks for 2nd-order SPT phases in $d=3$ are simply 2d $G_0$-SPT phases, classified by $\gpc{3}{G_0}{U(1)}$ within the group cohomology framework.

Unlike 1d $G_0$-SPT phases which always form a finite Abelian group, 2d $G_0$-SPT phases can be an infinite Abelian group, as shown in the last line of TABLE \ref{tab:k=1,d=3}. Therefore to classify 2nd-order SPT phases in $d=3$, we need to compute $\gpc{1}{G_c^\ast}{\mbz}$ in addition to knowledge of $\gpc{1}{G_c^\ast}{\mbz_a}$ in (\ref{k=1,Za:1st})-(\ref{k=1,Za:last}). Below we summarize $\gpc{1}{G_c^\ast}{\mbz}$ for all axial point groups and magnetic point groups $G_c$:
\bea
&\notag (C_n)^\ast\simeq Z_n,~~(S_{2n}^\bst)^\ast\simeq Z_{2n},\\
\label{k=1,Z:1st}&\gpc{1}{Z_n}{\mbz}=\mbz_{1};\\
&\notag (C_{n,\text{v}})^\ast\simeq(C_n\rtimes Z_2^{\bsm_\text{v}\bsm_\text{h}\bst})^\ast\simeq Z_n\rtimes Z_2^\bst,\\
&\gpc{1}{Z_n\rtimes Z_2^\bst}{\mbz}=\mbz_{2};\\
&\notag(C_{2n}^\bst)^\ast\simeq(S_{2n})^\ast\simeq Z_{2n}^\bst;\\
&\gpc{1}{Z_{2n}^\bst}{\mbz}=\mbz_{2};\\
&\notag(D_{n,\text{d}})^\ast\simeq(C_{2n}^\bst\rtimes Z_2^{\bsm_\text{v}})^\ast\simeq Z_{2n}^\bst\rtimes Z_2,\\
&\gpc{1}{Z_{2n}^\bst\rtimes Z_2}{\mbz}=\mbz_{2};\\
&\notag(D_n)^\ast\simeq(C_n\rtimes Z_2^{\bsm_\text{v}\bst})^\ast\simeq Z_{n}\rtimes Z_2,\\
&\gpc{1}{Z_n\rtimes Z_2}{\mbz}=\mbz_{1};\\
&\notag (C_{n,\text{h}})^\ast\simeq(C_n\times Z_2^\bsi)^\ast\simeq Z_n\times Z_2^\bst,\\
&\gpc{1}{Z_n\times Z_2^\bst}{\mbz}=\mbz_{2};\\
&\notag (D_{n,\text{h}})^\ast\simeq(Z_n\rtimes Z_2)\times Z_2^\bst,\\
&\gpc{1}{(Z_n\rtimes Z_2)\times Z_2^\bst}{\mbz}=\mbz_{2}.\label{k=1,Z:last}
\eea

Using relation (\ref{factorization}) and results (\ref{k=1,Za:1st})-(\ref{k=1,Za:last}), (\ref{k=1,Z:1st})-(\ref{k=1,Z:last}), we are able to compute $\gpc{1}{G_c^\ast}{\gpc{3}{G_0}{U(1)}}$ for various onsite symmetry $G_0$. The classification of 2nd-order SPT phases in $d=3$ is summarized in TABLE \ref{tab:k=1,d=3}.

It is known that there are certain 2d short-range-entangled (SRE) bosonic phases (without intrinsic topological order) exhibiting chiral edge states\cite{Kitaev2006,Lu2012a}, which are beyond the description of group cohomology classification. These SRE bosonic phases have an integer ($\mbz$) classification, generated by the $E_8$ state with a chiral central charge $c_-=8$. One can also build higher-order SPT phases out of the bosonic $E_8$ states, as highlighted by the red color in TABLE \ref{tab:k=1,d=3}.

\subsection{Examples}

One physical signature of 2nd-order SPT phases in $d=3$ is the existence of gapless states on certain 1d hinges of the system. Below we elucidate the procedure of constructing 2nd-order SPT phases in $d=3$ using the data of $\gpc{1}{G_c^\ast}{\gpc{3}{G_0}{U(1)}}$, based on the decorated domain wall construction where the building blocks are 2d $G_0$-SPT phases and bosonic $E_8$ states. We also show how this construction leads to gapless hinge states in 3d 2nd-order SPT phases.

\subsubsection{$G_c=C_{n}$}

\begin{figure}
\includegraphics[width=150pt]{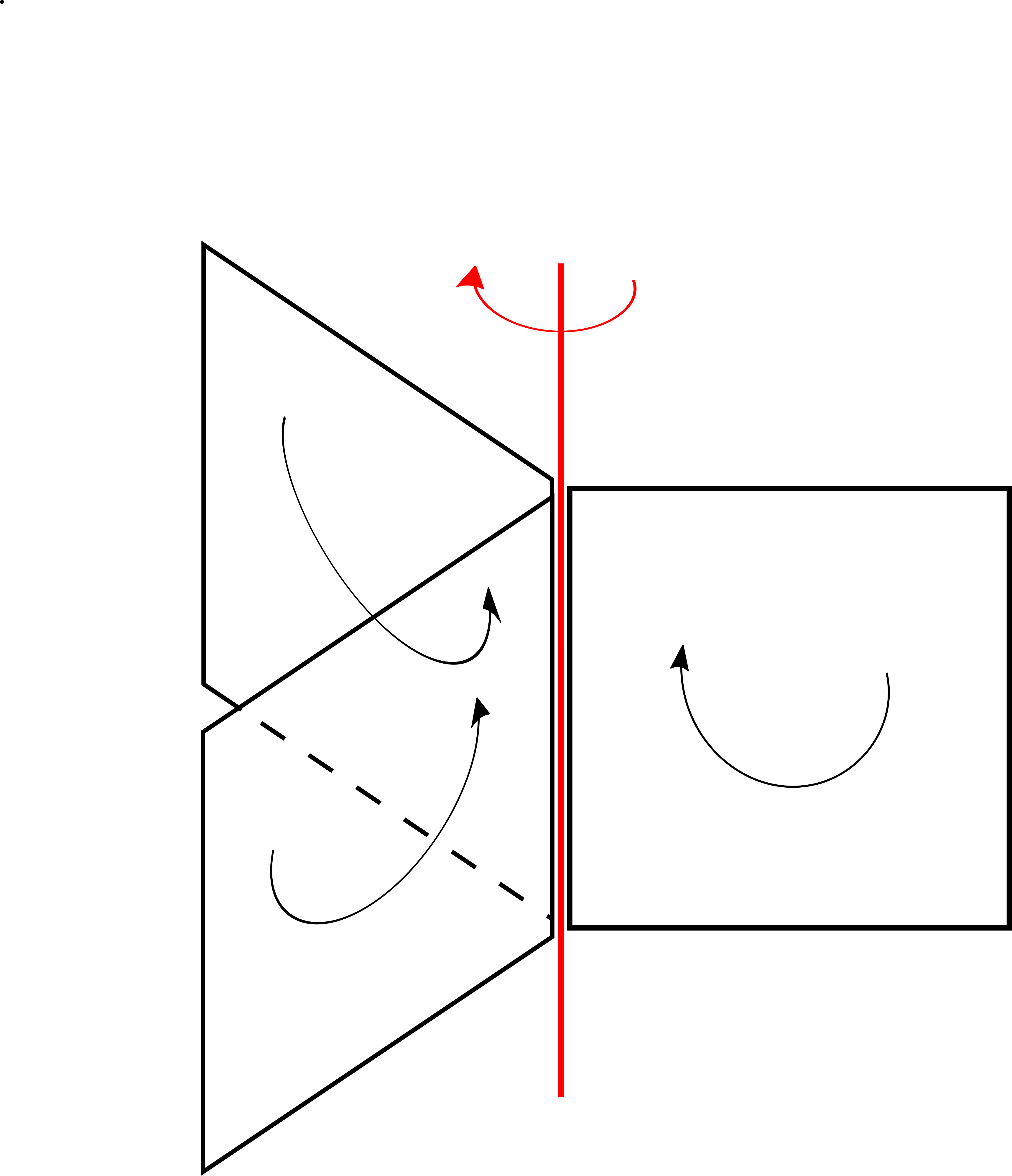}
\caption{2nd-order SPT phases in $d=3$ spatial dimensions, preserving $C_3$ rotational symmetry. Similar to $d=2$ case in FIG. \ref{fig:c3}, they can be constructed by assigning the same 2d $G_0$-SPT phases on each $C_3$ domain wall.}\label{fig:C3:3d}
\end{figure}

In the simplest case of $n$-fold rotational symmetry $G_c=C_n$, the 2nd-order SPT phases in 3d can be constructed by decorating each $C_n$ domain wall by the same 2d $G_0$-SPT phase, as illustrated in FIG. \ref{fig:C3:3d}. Similar to 2d cases discussed earlier, a gapped bulk provides strong constraints on the compatible 2d $G_0$-SPT phases, encoded in the following group cohomology formulae:
\bea
\gpc{1}{Z_n}{\mbz_a}=\mbz_{(n,a)}
\eea
and
\bea
\gpc{1}{Z_n}{\mbz}=\mbz_{1}
\eea
This means if the 2d $G_0$-SPT phases has an integer classification \ie $\gpc{3}{G_0}{U(1)}=\mbz$, none of these SPT phases are compatible to a gapped bulk when decorated on the $C_n$ domain walls. On the other hand, if the 2d $G_0$-SPT phases form a finite group such as $\gpc{3}{G_0}{U(1)}=\mbz_a$, only those with a topological index $\nu=0\mod\frac{a}{(n,a)}$ can lead to a gapped spectrum at the $C_n$ rotation center, indicated by (\ref{rep:Cn}) discussed earlier. Similar to $d=2$ cases in section \ref{sec:k=1,d=2}, these 2d $G_0$-SPT phases can be rotated together by an arbitrary angle around the $C_n$ axis. Notice that 1d gapless hinge modes are not always present in a finite system: they only appear when the hinge intersects with the plane of each 2d $G_0$-SPT phase.

\subsubsection{$G_c=C_{n,\text{v}}$}

Considering point group $G_c=C_{n,\text{v}}$ or magnetic point group $C_n\rtimes Z_2^{\bsm_\text{v}\bsm_\text{h}\bst}$, the construction of associated 2nd-order SPT phases in $d=3$ is completely in parallel to $d=2$ cases illustrated in FIG. \ref{fig:c3m}. Specifically, two types of 2d $G_0$-SPT phases are assigned to each mirror plane: one type (red lines in FIG. \ref{fig:c3m}) meeting at the $C_n$ rotation center corresponds to the linear representation $U_R$ of $n$-fold rotation generator $R$, the other type (green lines in FIG. \ref{fig:c3m}) on each mirror planet corresponds to linear representation $U_{\bsm_\text{v}}$ of mirror operation $\bsm_\text{v}$. They are constrained by the following compatibility conditions for a gapped bulk. When $\gpc{3}{G_0}{U(1)}$ \ie the classification of 2d $G_0$-SPT phases is a finite group, we have
\bea
\gpc{1}{Z_n\rtimes Z_2^\bst}{\mbz_a}=\mbz_{(n,a)}\times\mbz_{(2,a)}
\eea
where $U_R\in\mbz_{(n,a)}$ and $U_{\bsm_\text{v}}\in\mbz_{(2,a)}$, the same as discussed in section \ref{sec:k=1,d=2} for $d=2$ case.

Meanwhile if the classification of 2d $G_0$-SPT phases is an infinite group labeled \eg by an integer topological index $\nu\in\mbz$, we have
\bea
\gpc{1}{Z_n\rtimes Z_2^\bst}{\mbz}=\mbz_{2}
\eea
where $U_R\equiv u_0\in\mbz_1$ and $U_{\bsm_\text{v}}=u_{\nu_{\bsm_\text{v}}\mod2}\in\mbz_2$. Physically, for the $C_n$ rotation center to be gapped, one can only assign a trivial 2d phase on each $C_n$ domain wall, corresponding to the trivial representation $U_R\equiv u_0$. On the other hand, each mirror plane can be decorated by any 2d $G_0$-SPT phase with topological index $\nu_{\bsm_\text{v}}$. The 2nd-order SPT phases is only characterized by the parity of topological index $\nu_{\bsm_\text{v}}\mod2$, since a pair of the same 2d $G_0$-SPT phases can always be merged onto the mirror plane without closing the bulk gap.

As shown in FIG. \ref{fig:c3m}, the gapless hinge states will appear in a finite system as long as the hinge lies within a mirror plane. The gapless 1d modes on the two opposite hinges of the same mirror plane are generally different from each other, as illustrated by the green hinges versus green-plus-red hinges in FIG. \ref{fig:c3m}.

\begin{figure}
\includegraphics[width=150pt]{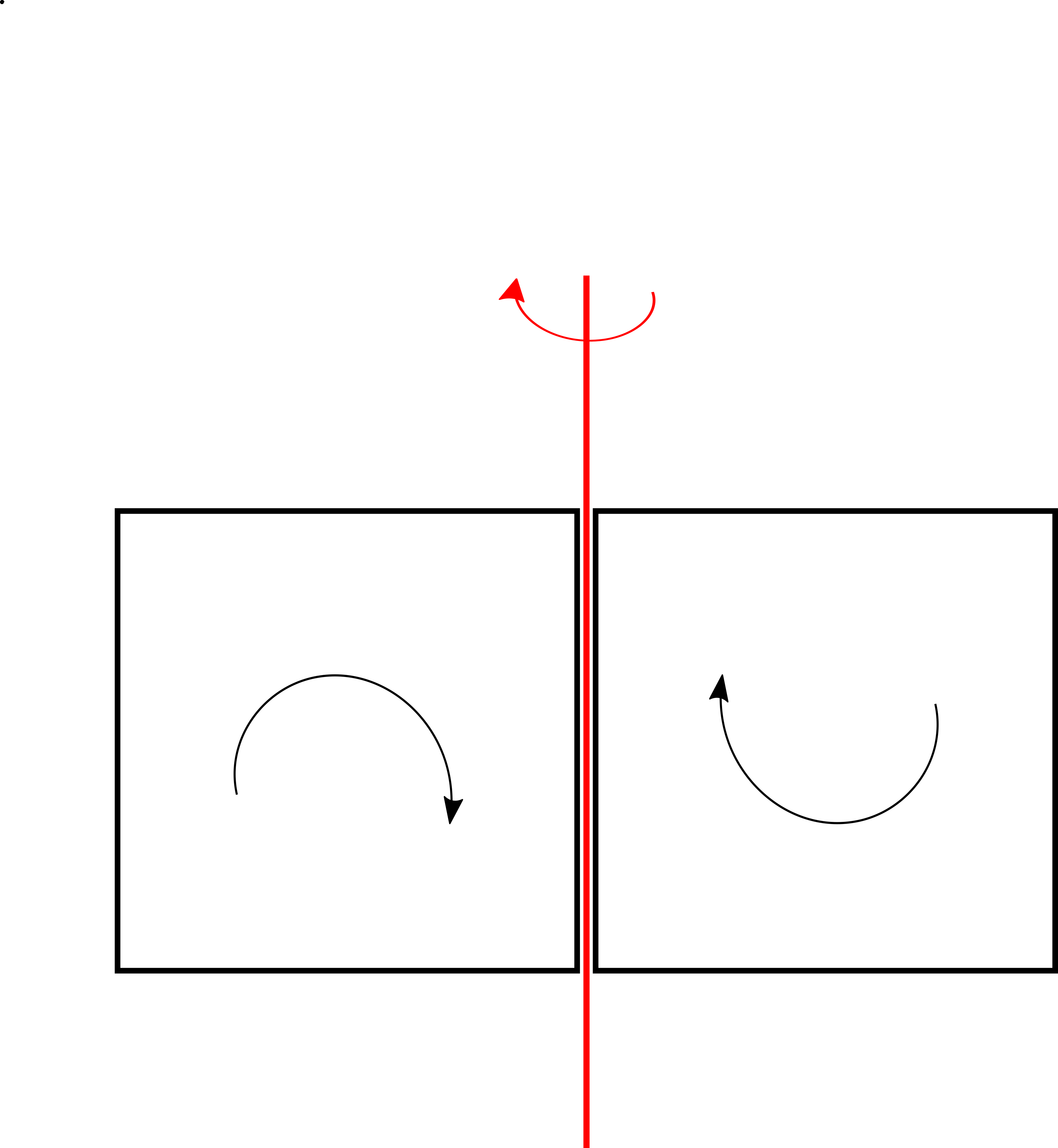}
\caption{2nd-order SPT phases with point group symmetry $S_2$ or magnetic point group $C_2^\bst=\{(R\cdot\bst)^i|i=0,1\}$.}
\label{fig:S2n:3d}
\end{figure}

\subsubsection{$G_c=S_{2n}$ or $G_c=C_{2n}^\bst$}

Point group $S_{2n}$ is generated by a $\frac{\pi}{n}$ rotation $R$ along $\hat z$-axis followed by a mirror $\bsm_\text{h}$ w.r.t. to [001] plane:
\bea
S_{2n}=\{\bss^i|1\leq i\leq 2n\},~~\bss=R\cdot\bsm_\text{h}\Longrightarrow\bss^{2n}=1.
\eea
Operation $\bss$ is usually referred to as an improper rotation or a rotoreflection. For both point group $S_{2n}$ and magnetic point group $C_{2n}^\bst$ defined below
\bea
C_{2n}^\bst\equiv\{(R\cdot\bst)^i|1\leq i\leq 2n\}
\eea
they share the same classification for 2nd-order SPT phases since
\bea
(S_{2n})^\ast\simeq (C_{2n}^\bst)^\ast\simeq Z_{2n}^\bst
\eea
where $Z_{2n}^\bst$ is generated by an anti-unitary operator of rank $2n$. The following group cohomology formulae determine the classification of 2nd-order SPT phases with $S_{2n}$ or $C_{2n}^\bst$ symmetry:
\bea\label{gpc:1,S2n,Za}
\gpc{1}{Z_{2n}^\bst}{\mbz_a}=\mbz_{(2,a)}
\eea
and
\bea\label{gpc:1,S2n,Z}
\gpc{1}{Z_{2n}^\bst}{\mbz}=\mbz_{2}
\eea
They are determined by solving the following conditions for linear representation $U_\bss\in\mbz_a,\mbz$:
\bea
U_\bss U_\bss^\ast=1.
\eea
They can be understood similar to the mirror symmetry $\bsm_\text{v}$ in the $G_c=C_{n,\text{v}}$ case, where we have $U_\bss=u_{\nu_{\bss}\mod2}$ and the topological index $\nu_{\bss}$ of the 2d $G_0$-SPT phase is only well defined modulo 2. To construct these $S_{2n}$-symmetric 2nd-order SPT phases, we decorate each $\bss$ domain wall by a 2d $G_0$-SPT phase with topological index $\nu_\bss\mod2$, in a staggered fashion as shown in FIG. \ref{fig:S2n:3d}. Again we can always merge two identical $G_0$-SPT phases into each $\bss$ domain wall without closing the bulk gap, which will change the topological index of 2d $G_0$ SPT phase on this $\bss$ domain wall by an even integer. This physically explains why the topological index for the 2d $G_0$-SPT phase $\nu_\bss$ decorated on each $\bss$ domain wall is only defined modulo 2, manifested in the $\mbz_{(2,a)}$ and $\mbz_2$ classification in (\ref{gpc:1,S2n,Za}) and (\ref{gpc:1,S2n,Z}).

\begin{figure}
\includegraphics[width=150pt]{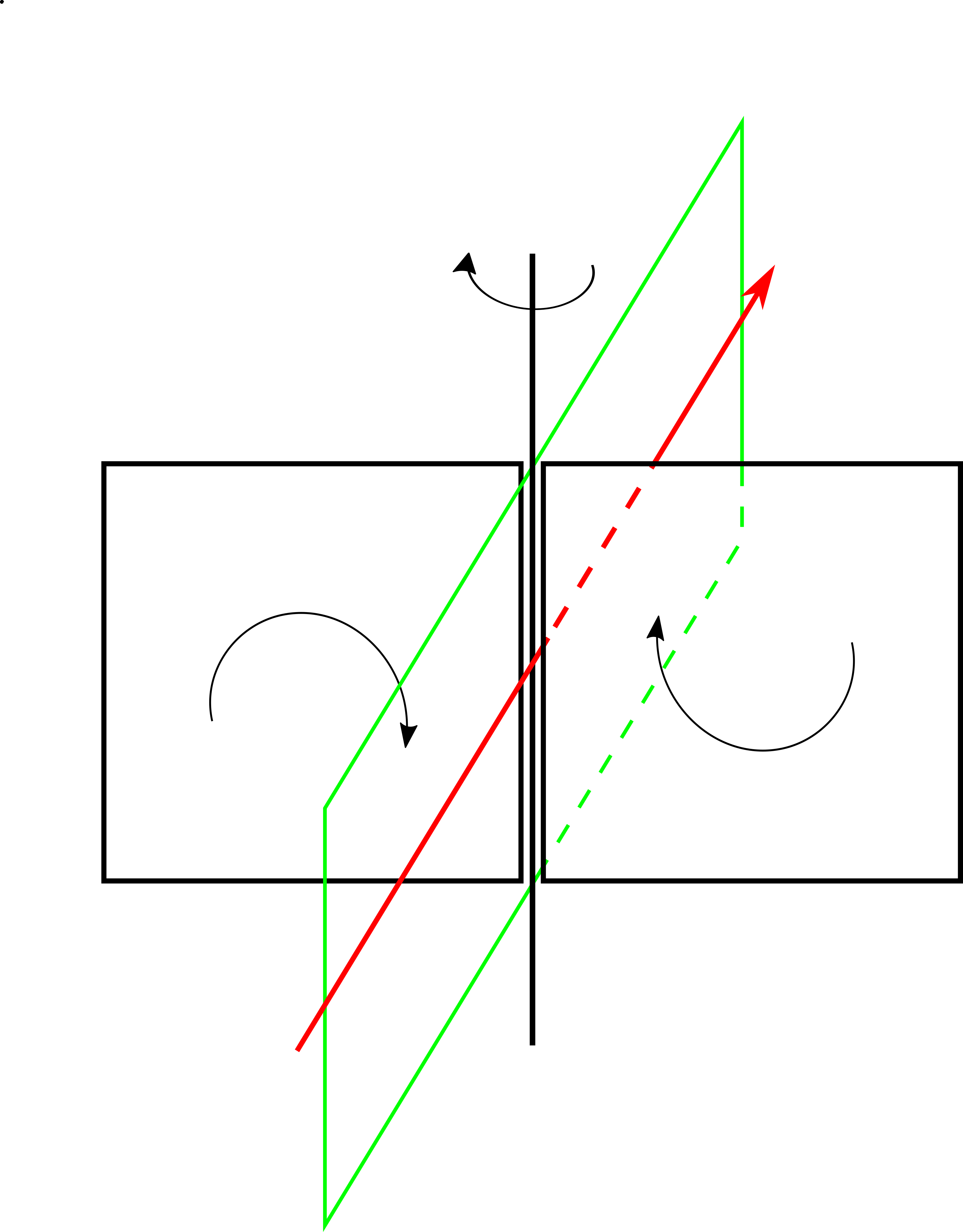}
\caption{2nd-order SPT phases with point group symmetry $D_{n,\text{d}}$. There is a 2-fold rotoreflection axis along $\hat z$-axis and a 2-fold in-plane axis (colored red). 2d $G_0$-SPT phases with topological index $\bss$ are decorated on each mirror plane, while 2d $G_0$-SPT phases with index $\nu_{R_2}$ are decorated on vertical planes crossing each $R_2$ axis.}
\label{fig:3d:Dnd}
\end{figure}

\subsubsection{$G_c=D_{n,\text{d}}\equiv S_{2n}\rtimes Z_2^{\bsm_\text{v}}$ or $C_{2n}^\bst\rtimes Z_2^{\bsm_\text{v}}$}

Point group $D_{n,\text{d}}$ is generated by $2n$-fold rotoreflection $\bss=R\cdot\bsm_\text{h}$ around $\hat z$-axis as discussed earlier in $G_c=S_{2n}$ case, and a mirror plane $\bsm_\text{v}$ parallel to $\hat z$-axis. The group $D_{n,\text{d}}$ can be summarized as
\bea
D_{n,\text{d}}=\{\bss^{i_s}(R_2)^{i_2}|1\leq i_s\leq 2n,1\leq i_2\leq 2.\}
\eea
where we defined $R_2\equiv \bss\cdot\bsm_\text{v}$ as a 2-fold rotation along an in-plane axis (colored red in FIG. \ref{fig:3d:Dnd}), so that $\bss^{2n}=(\bss\cdot\bsm_\text{v})^2=1$. The linear representation $\gpc{1}{G_c^\ast}{\gpc{3}{G_0}{U(1)}}=\{U_\bss,U_{R_2}\in\gpc{3}{G_0}{U(1)}\}$ must satisfy the following algebraic conditions:
\bea
&U_\bss U_\bss^\ast=1,\\
&(U_{R_2})^2=1,\\
&U_{R_2}U_\bss(U_{R_2} U_\bss)^\ast=1.
\eea
If 2d $G_0$-SPT phases has a finite classification such as $\gpc{3}{G_0}{U(1)}=\mbz_a$, the linear representations are classified as
\bea
\gpc{1}{Z_{2n}^\bst\rtimes Z_2}{\mbz_a}=\mbz_{(2,a)}^2
\eea
given by
\bea
&U_\bss=e^{\frac{2\pi\imth}{a}\nu_\bss},~~~\nu_\bss\simeq\nu_\bss+2,\\
&U_{R_2}=e^{\frac{2\pi\imth}{a}\nu_{R_2}},~~~2\nu_{R_2}=0\mod a.
\eea
where $\nu_\bss$ and $\nu_{R_2}$ are the topological indices of 2d $G_0$-SPT phases decorated on $\bss$ and $R_2$ domain walls respectively.

If 2d $G_0$-SPT phases has an infinite classification such as $\gpc{3}{G_0}{U(1)}=\mbz$, we have
\bea
\gpc{1}{Z_{2n}^\bst\rtimes Z_2}{\mbz}=\mbz_{2}
\eea
where
\bea
\nu_\bss=0,1\simeq\nu_\bss+2,~~~\nu_{R_2}=0.
\eea
Physically, the 2d $G_0$-SPT phases decorated on each $\bss$ domain wall (chosen to lie within a mirror plane) have topological indices $\nu_\bss=0,1$ defined modulo 2, for the same reason described previously in $G_c=S_{2n}$ case. They are illustrated by black color in FIG. \ref{fig:3d:Dnd}. On the other hand, the topological index $\nu_{R_2}\simeq-\nu_{R_2}$ decorated on each $R_2$ domain wall must be non-chiral, and hence must be trivial when $\gpc{3}{G_0}{U(1)}=\mbz$. These $\nu_{R_2}$-indexed 2d $G_0$-SPT phases are decorated on vertical planes parallel to each $R_2$ axis, as illustrated by the green plane in FIG. \ref{fig:3d:Dnd}.

Clearly the hinges of a finite system can host 1d gapless modes for these 2nd-order SPT phases, if the hinge lies within a mirror plane or a vertical plane containing one 2-fold axis. Generally the gapless modes on these two types of hinges will be different.

\begin{figure}
\includegraphics[width=150pt]{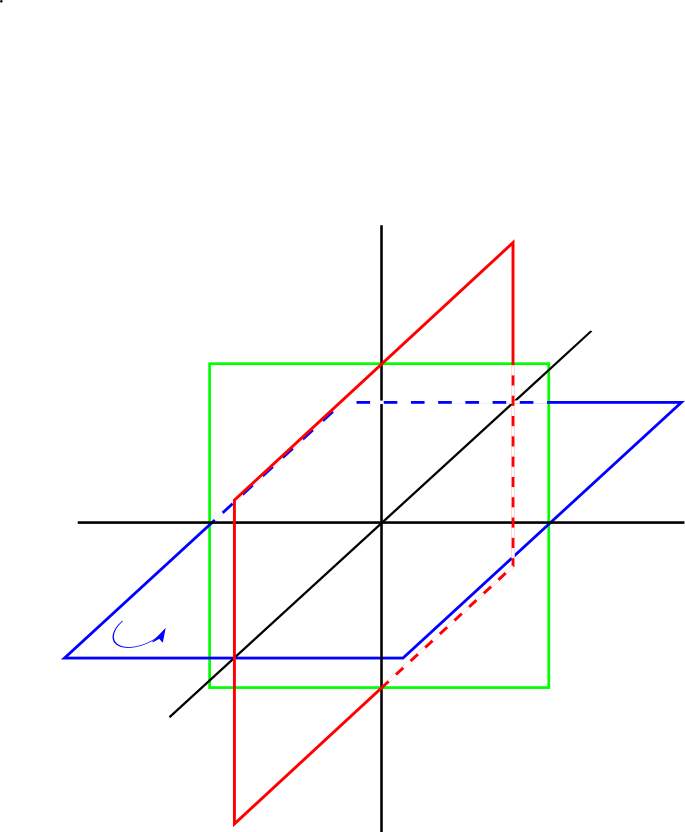}
\caption{2nd-order SPT phases with $D_{n,\text{h}}$ point group symmetry. 2d $G_0$-SPT phases with topological indices $\nu_{R_z}$, $\nu_{R_x}$ and $\nu_{\bsm_\text{h}}$ are assigned to red, green and blue mirror planes respectively.}
\label{fig:3d:Dnh}
\end{figure}

\subsubsection{$G_c=D_{n,\text{h}}\equiv C_{n,\text{v}}\times Z_2^{\bsm_\text{h}}$}

Finally we consider the following point group
\bea
D_{n,\text{h}}=\{(R_z)^{i_n}(R_x)^{i_2}(\bsm_\text{h})^{i_\bsm}|i_n\in Z_n,i_2,i_\bsm\in Z_2\}.
\eea
As shown in FIG. \ref{fig:3d:Dnh}, it is generated by $n$-fold rotation $R_z$ along $\hat z$-axis, 2-fold rotation $R_x$ along $\hat x$-axis and mirror $\bsm_\text{h}$ w.r.t to the $x$-$y$ (or [001]) plane (colored in blue in FIG . \ref{fig:3d:Dnh}). Its linear representation $\{U_{R_x},U_{R_z},U_{\bsm_\text{h}}\in\gpc{3}{G_0}{U(1)}\}\in\gpc{1}{D_{n,\text{h}}^\ast}{\gpc{3}{G_0}{U(1)}}$ satisfies the following conditions:
\bea
&(U_{R_z})^n=(U_{R_x})^2=(U_{R_x}U_{R_z})^2=1,\\
&U_{\bsm_\text{h}}U_{\bsm_\text{h}}^\ast=1,\\
&U_{R_z}U_{\bsm_\text{h}}(U_{R_z}U_{\bsm_\text{h}})^\ast=1,\\
&U_{R_x}U_{\bsm_\text{h}}(U_{R_x}U_{\bsm_\text{h}})^\ast=1.
\eea
When 2d $G_0$-SPT phases have a finite classification, \eg $\gpc{3}{G_0}{U(1)}=\mbz_a$ we have
\bea
\gpc{1}{(Z_n\rtimes Z_2)\times Z_2^\bst}{\mbz_a}=\mbz_{(n,a,2)}\times\mbz_{(2,a)}^2
\eea
where
\bea
&U_{R_z}=e^{\frac{2\pi\imth}{a}\nu_{R_z}}\in\mbz_{(n,a,2)},~n\nu_{R_z}=2\nu_{R_z}=0\mod a,~~~\\
&U_{R_x}=e^{\frac{2\pi\imth}{a}\nu_{R_x}}\in\mbz_{(a,2)},~~~2\nu_{R_x}=0\mod a,\\
&U_{\bsm_\text{h}}=e^{\frac{2\pi\imth}{a}\nu_{\bsm_\text{h}}}\in\mbz_{(a,2)},~~~\nu_{\bsm_\text{h}}=0,1\mod2.
\eea
Meanwhile if 2d $G_0$-SPT phases have an infinite classification, \eg $\gpc{3}{G_0}{U(1)}=\mbz$ we have
\bea
\gpc{1}{(Z_n\rtimes Z_2)\times Z_2^\bst}{\mbz}=\mbz_{2}
\eea
where
\bea
\nu_{R_z}=\nu_{R_x}\equiv0,~~~\nu_{\bsm_\text{h}}=0,1\mod2\in\mbz_2.
\eea
Physically, both $R_z$ and $R_x$ domain walls must be decorated with non-chiral 2d $G_0$-SPT phases due to the mirror symmetry $\bsm_\text{h}$, as denoted by red ($\nu_{R_z}$) and green ($\nu_{R_x}$) in FIG. \ref{fig:3d:Dnh}. Meanwhile, the $\bsm_\text{h}$ mirror plane can be decorated with a (possibly chiral) 2d $G_0$-SPT phases with index $\nu_{\bsm_\text{h}}$, as denoted by blue in FIG. \ref{fig:3d:Dnh}.

From the above construction, we can see that there are three types of hinges hosting different gapless 1d modes in the system: (i) 1d hinge modes lying in the mirror $\bsm_\text{h}$ plane (colored in blue) has topological index $\nu_{\bsm_\text{h}}$; (ii) 1d hinge modes lying in vertical mirror $R_x\cdot\bsm_\text{h}$ plane (colored in green) has index $\nu_{R_x}$; (iii) 1d hinge modes lying in vertical mirror $R_z\cdot R_x\cdot\bsm_\text{h}$ plane (colored in red) has index $\nu_{R_z}$.

\begin{table*}[tb]
\centering
\begin{tabular} {|c||c|c|c|c|}
\hline
$k=2,d=3$&\multicolumn{4}{c|}{Onsite symmetry $G_0$}\\
\hline
&$Z_2^\bst,~SO(3)$&$SO(3)\times Z_2^\bst$ &$Z_a\times Z_b$&$Z_a\times Z_2^\bst$\\
Crystalline &or $U(1)\rtimes Z_2$&or $U(1)\times Z_2^\bst$&&\\
symmetry $G_c$&&&&\\
\hline
$C_n$&$\mbz_{(2,n)}$&$\mbz_{(2,n)}^2$&$\mbz_{(n,a,b)}$&$\mbz_{(2,n)}\times\mbz_{(a,2,n)}$\\
\hline
$C_{n,\text{v}}$&$\mbz_{(n,2)}\times\color{green}{\mbz_{(n,2)}\times\mbz_2}$&$\mbz_{(n,2)}^2\times\color{green}{\mbz_{(n,2)}^2\times\mbz_2^2}$&$\mbz_{(n,a,b)}\times\color{green}{\mbz_{(n,a,b,2)}\times\mbz_{(2,a,b)}}$&$\mbz_{(n,2)}\times\mbz_{(n,a,2)}\times\color{green}{\mbz_{(n,a,2)}\times\mbz_{(a,2)}\times\mbz_{(n,2)}\times\mbz_2}$\\
\hline
$C_{n,\text{h}}$&$\mbz_{(n,2)}\times\mbz_2\times\color{green}{\mbz_2}$&$\mbz_{(n,2)}^2\times\mbz_2^2\times\color{green}{\mbz_2^2}$&$\mbz_{(n,a,b,2)}\times\mbz_{(a,b,2)}\times\color{green}{\mbz_{(a,b,2)}}$&$\mbz_{(n,2)}\times\mbz_2\times\mbz_{(n,a,2)}\times\mbz_{(a,2)}\times\color{green}{\mbz_2\times\mbz_{(a,2)}}$\\
\hline
$D_{n}$&$\mbz_{(n,2)}\times\mbz_2^2$&$\mbz_{(n,2)}^2\times\mbz_2^4$&$\mbz_{(n,a,b,2)}\times\mbz_{(2,a,b)}^2$&$\mbz_{(n,2)}\times\mbz_2^2\times\mbz_{(n,a,2)}\times\mbz_{(a,2)}^2$\\
\hline
$C_{n}\rtimes Z_2^{\bsm_\text{v}\cdot\bst}$&$\mbz_{(n,2)}\times\color{green}{\mbz_2^2}$&$\mbz_{(n,2)}^2\times\color{green}{\mbz_2^4}$&$\mbz_{(n,a,b,2)}\times\color{green}{\mbz_{(a,b,2)}^2}$&$\mbz_{(n,2)}\times\mbz_{(n,a,2)}\times\color{green}{\mbz_2^2\times\mbz_{(a,2)}^2}$\\
\hline

$D_{n,\text{h}}$&$\mbz_{(n,2)}^2\times\mbz_2\times\color{green}{\mbz_2^3}$&$\mbz_{(n,2)}^4\times\mbz_2^2\times\color{green}{\mbz_2^6}$&$\mbz_{(n,a,b,2)}^2\times\mbz_{(a,b,2)}\times\color{green}{\mbz_{(a,b,2)}^3}$&$\mbz_{(n,2)}^2\times\mbz_2\times\mbz_{(n,a,2)}^2\times\mbz_{(a,2)}\times\color{green}{\mbz_2^3\times\mbz_{(a,2)}^3}$\\
\hline

$C_{2n}^\bst$ or $S_{2n}$&$\mbz_2$&$\mbz_2^2$&$\mbz_{(2,a,b)}$&$\mbz_2\times\mbz_{(a,2)}$\\
\hline
$C_{2n}^\bst\rtimes Z_2^{\bsm_\text{v}}$&$\mbz_2\times\color{green}{\mbz_2^2}$&$\mbz_2^2\times\color{green}{\mbz_2^4}$&$\mbz_{(a,b,2)}\times\color{green}{\mbz_{(a,b,2)}^2}$&$\mbz_2\times\mbz_{(a,2)}\times\color{green}{\mbz_2^2\times\mbz_{(a,2)}^2}$\\
\hline
$D_{n,\text{d}}$&$\mbz_2^2\times\color{green}{\mbz_2}$&$\mbz_2^4\times\color{green}{\mbz_2^2}$&$\mbz_{(a,b,2)}^2\times\color{green}{\mbz_{(a,b,2)}}$&$\mbz_2^2\times\mbz_{(a,2)}^2\times\color{green}{\mbz_2\times\mbz_{(a,2)}}$\\
\hline
$S_{2n}^\bst$&$\mbz_{2}$&$\mbz_{2}^2$&$\mbz_{(2n,a,b)}$&$\mbz_{2}\times\mbz_{(a,2)}$\\
\hline
$T\simeq A_4$&$\mbz_{2}$&$\mbz_{2}^2$&$\mbz_{(3,a,b)}\times \mbz_{(2,a,b)}$&$\mbz_{2}\times\mbz_{(a,2)}$\\
\hline
$T_h=T\times Z_2^{\bsi}$&$\mbz_2^2\times\color{green}{\mbz_2}$&$\mbz_2^4\times\color{green}{\mbz_2^2}$
&$\mbz_{(2,a,b)}^2\times\color{green}{\mbz_{2,a,b}^2}$&$\mbz_2^2\times\mbz_{(2,a)}^2\times\color{green}{\mbz_2\times\mbz_{(2,a)}}$\\
%\hline
%$T_d$&&&&&\\
%\hline
%$O\simeq S_4$&&&&&\\
%\hline
%$O_h=O\times Z_2^{\bsi}$&&&&&\\
\hline
\hline
\color{blue}{$\gpc{2}{G_0}{U(1)}$}&\color{blue}{$\mbz_2$}&\color{blue}{$\mbz_2^2$}&\color{blue}{$\mbz_{(a,b)}$}&\color{blue}{$\mbz_2\times\mbz_{(a,2)}$}\\
\hline
\end{tabular}
\caption{3rd-order bosonic SPT phases ($k=1$) in $d=2$ spatial dimensions, protected by symmetry group $G_s=G_c\times G_0$ where $G_c$ and $G_0$ represent the crystalline and onsite symmetry group respectively. The general classification is given by projective representation $\gpc{2}{G_c^\ast}{\gpc{2}{G_0}{U(1)}}$ as shown in (\ref{class:3rd order}) with $d=3$. Through a dimensional reduction procedure they are all constructed from $G_0$-SPT phases in $d=1$ dimension, classified by $\gpc{2}{G_0}{U(1)}$ in the last line (blue). To be contrasted with the strong 3rd-order SPT phases in black, the weak crystalline SPT phases included in $\gpc{2}{G_c^\ast}{\gpc{2}{G_0}{U(1)}}$ are colored in green.}
\label{tab:k=2,d=3}
\end{table*}

\section{3rd-order SPT phases in three dimensions}\label{sec:k=2,d=3}

The 3rd-order SPT phases in $d=3$ are classified by 2nd group cohomology $\gpc{2}{G_c^\ast}{\gpc{2}{G_0}{U(1)}}$ whose coefficients take value in $\gpc{2}{G_0}{U(1)}$. Physically they can be constructed by stacking 1d $G_0$-SPT phases \ie elements of $\gpc{2}{G_0}{U(1)}$, in a way which preserves (magnetic) crystalline symmetry $G_c$. Specifically as previously discussed in the decorated domain wall picture in section \ref{sec:ddw}, they can all be built by decorating 1d intersections of domain walls with 1d $G_0$-SPT phases. Below we first classify these 3rd-order SPT phases in $d=3$ dimensions, for various onsite symmetry $G_0$ and (magnetic) point group $G_c$. Then we illustrate how to explicitly construct these states in a few examples.

\subsection{Classification}

To compute the group cohomology $\gpc{2}{G_c^\ast}{\gpc{2}{G_0}{U(1)}}$ for 3rd-order SPT phases, we first notice that the classification of 1d $G_0$-SPT phases always form a finite discrete Abelian group, which are products of the cyclic group $\mbz_a$ for a finite $a\in\mbz$. Therefore according to relation (\ref{factorization}), we only need to know $\gpc{2}{G_c^\ast}{\mbz_a}$ in order to compute $\gpc{2}{G_c^\ast}{\gpc{2}{G_0}{U(1)}}$.

Below we list the results for various point groups and magnetic point groups $G_c$:
\bea
&\notag (C_n)^\ast\simeq Z_n,~~(S_{2n}^\bst)^\ast\simeq Z_{2n},\\
\label{k=2,Za:1st}&\gpc{2}{Z_n}{\mbz_a}=\mbz_{(n,a)};\\
&\notag (C_{n,\text{v}})^\ast\simeq(C_n\rtimes Z_2^{\bsm_\text{v}\bsm_\text{h}\bst})^\ast\simeq Z_n\rtimes Z_2^\bst,\\
\label{k=2,d=3:magnetic Cnv}&\gpc{2}{(C_n\rtimes Z_2^{\bsm_\text{v}\bsm_\text{h}\bst})^\ast}{\mbz_a}=\mbz_{(n,a,2)}\times\mbz_{(n,a)}\times\mbz_{(2,a)};~~~\\
\label{k=2,d=3:Cnv}&\gpc{2}{(C_{n,\text{v}})^\ast}{\mbz_a}=\mbz_{(n,a)}\times\color{green}{\mbz_{(n,a,2)}\times\mbz_{(2,a)}};\\
&\notag(C_{2n}^\bst)^\ast\simeq(S_{2n})^\ast\simeq Z_{2n}^\bst;\\
&\gpc{2}{Z_{2n}^\bst}{\mbz_a}=\mbz_{(2,a)};\\
&\notag(D_{n,\text{d}})^\ast\simeq(C_{2n}^\bst\rtimes Z_2^{\bsm_\text{v}})^\ast\simeq Z_{2n}^\bst\rtimes Z_2,\\
&\gpc{2}{(D_{n,\text{d}})^\ast}{\mbz_a}=\mbz_{(2,a)}^2\times\color{green}{\mbz_{(2,a)}};\\
&\gpc{2}{(C_{2n}^\bst\rtimes Z_2^{\bsm_\text{v}})^\ast}{\mbz_a}=\mbz_{(2,a)}\times\color{green}{\mbz_{(2,a)}^2};\\
&\notag(D_n)^\ast\simeq(C_n\rtimes Z_2^{\bsm_\text{v}\bst})^\ast\simeq Z_{n}\rtimes Z_2,\\
&\gpc{2}{(D_n)^\ast}{\mbz_a}=\mbz_{(n,a,2)}\times\mbz_{(a,2)}^2;\\
&\gpc{2}{(C_n\rtimes Z_2^{\bsm_\text{v}\bst})^\ast}{\mbz_a}=\mbz_{(n,a,2)}\times\color{green}{\mbz_{(a,2)}^2};\\
&\notag (C_{n,\text{h}})^\ast\simeq(C_n\times Z_2^\bsi)^\ast\simeq Z_n\times Z_2^\bst,\\
&\gpc{2}{Z_n\times Z_2^\bst}{\mbz_a}=\mbz_{(n,a,2)}\times\mbz_{(2,a)}\times\color{green}{\mbz_{(2,a)}};\\
&\notag (D_{n,\text{h}})^\ast\simeq(Z_n\rtimes Z_2)\times Z_2^\bst,\\
&\gpc{2}{(D_{n,\text{h}})^\ast}{\mbz_a}=\mbz_{(n,a,2)}^2\times\mbz_{(2,a)}\times\color{green}{\mbz_{(2,a)}^3};\\
&\notag T^\ast\simeq A_4=(Z_2\times Z_2)\rtimes Z_3,\\
&\gpc{2}{A_4}{\mbz_a}=\mbz_{(3,a)}\times\mbz_{(2,a)};\\
&\notag (T_\text{h})^\ast\simeq A_4\times Z_2^\bst,\\
&\gpc{2}{A_4\times Z_2^\bst}{\mbz_a}=\mbz_{(2,a)}^2\times\color{green}{\mbz_{(2,a)}}.
%&\gpc{2}{O^\ast\simeq S_4}{\mbz_a}=\mbz_{(2,a)}^2;???\\
%&\gpc{2}{T_\text{d}^\ast\simeq A_4\rtimes Z_2^\bst}{\mbz_a}=\mbz_{(3,a)}\times\mbz_{(2,a)}^3;???\\
%&\gpc{2}{O_\text{h}^\ast\simeq S_4\times Z_2^\bst}{\mbz_a}=\mbz_{(2,a)}^5.???
\label{k=2,Za:last}
\eea
They are projective representations of group $G_c^\ast$ with coefficients valued in $\mbz_a$. Using these results and relation (\ref{factorization}), we obtain the classification of 3rd-order SPT phases for these (magnetic) point groups $G_c$ and various onsite symmetry $G_0$, as summarized in TABLE \ref{tab:k=2,d=3}.

\subsection{Examples}

\subsubsection{$G_c=C_n$}\label{sec:3ospt:Cn}

First we consider point group $C_n$, generated by rotation $R$ along \eg $\hat z$-axis. Its 2nd group cohomology is classified by
\bea\label{gpc:2,Cn,Za}
\gpc{2}{Z_n}{\mbz_a}=\mbz_{(n,a)}
\eea
which can be understood as follows. As described in section \ref{sec:ddw}, the 2nd group cohomology $\gpc{2}{Z_n}{\mbz_a}$ are projective representations $\{\omega(g,h)\in\mbz_a|g,h\in Z_n\}$ valued in $\mbz_a=\{e^{\frac{2\pi\imth}{a}\nu}|\nu\in\mbz\}$, defined below
\bea
&U_g\cdot U^{s(g)}_h=\omega(g,h)U_{gh},\\
&\omega(g,h)\omega(gh,k)=\omega(g,hk)\omega^{s(g)}(h,k).
\eea
For our group $(C_n)^\ast\simeq Z_n$ with $R^n=1$, we have
\bea
(U_R)^n=\omega_{C_n}\cdot1,~~~\omega_{C_n}=\prod_{i=1}^{n-1}\omega(R,R^i)=e^{\frac{2\pi\imth}{a}\nu_{C_n}}\in\mbz_a.
\eea
since $U_1\equiv1$. Notice that we can always redefine the symmetry operation by an extra phase factor valued in $\mbz_a$
\bea
U_R\rightarrow e^{\frac{2\pi\imth}{a}}U_R
\eea
which leads to equivalence relation
\bea\label{gauge:Cn}
\nu_{C_n}\simeq\nu_{C_n}+n\mod a\in\mbz_{(n,a)}.
\eea
where $(n,a)$ is the greatest common divisor of integers $n$ and $a$. This leads to the 2nd group cohomology formula (\ref{gpc:2,Cn,Za}).

Physically, as argued in section \ref{sec:ddw}, we decorate each $C_n$ rotational axis by a 1d $G_0$-SPT phase with topological index $\nu_{C_n}$. Notice that we can always merge $n$ copies of the same 1d $G_0$-SPT phases into the rotational axis in a $C_n$-symmetric manner, without closing the bulk gap. This physically explains the equivalence relation (\ref{gauge:Cn}).

Unlike 2nd-order SPT phases in $d=3$ which hosts gapless 1d hinge states, 3rd-order SPT phases in $d=3$ only supports gapless zero modes at the corners of certain finite systems, which carry projective representation of onsite symmetry $G_0$. For example here for point group $G_c=C_n$ or magnetic point group $G_c=S_{2n}^\bst=\{(\bss\bst)^i|1\leq i\leq2n\}$, there will be protected zero modes at every corner of the finite system lying on a $C_n$ rotatonal axis.

\begin{figure}
\includegraphics[width=150pt]{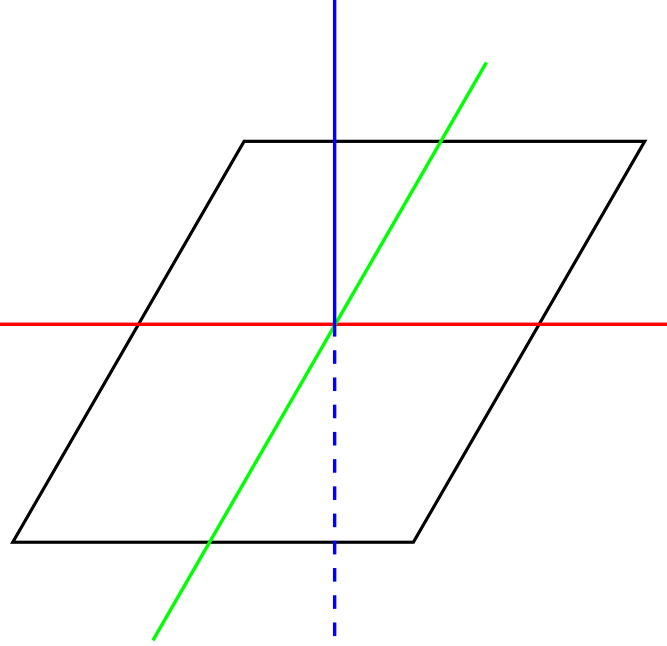}
\caption{3rd-order SPT phases with magnetic point group symmetry $C_n\rtimes Z_2^{\bsm_\text{v}\bsm_\text{h}\bst}$ with $n=2$. The topological indices of 1d $G_0$-SPT phases are labeled by $\nu_{R}$ (colored blue, along $n$-fold vertical rotation axis $R$), $\nu_\bsc$ (colored red, along 2-fold horizontal magnetic rotation axis $\bsc$) and $\nu_{R\bsc}$ (colored green, along 2-fold horizontal magnetic rotation axis $R\cdot\bsc$).}
\label{fig:k=2:Cnv}
\end{figure}

\subsubsection{$G_c=C_n\rtimes Z_2^{\bsm_\text{v}\bsm_\text{h}\bst}$ and $C_{n,\text{v}}$}\label{sec:3ospt:Cnv}

Next we consider magnetic point group $C_n\rtimes Z_2^{\bsm_\text{v}\bsm_\text{h}\bst}$ and point group $C_{n,\text{v}}$, which share the same
\bea
G_c^\ast\simeq Z_2\rtimes Z_2^\bst.
\eea
Take magnetic point group $G_c=C_n\rtimes Z_2^{\bsm_\text{v}\bsm_\text{h}\bst}$ for example: generated by $n$-fold vertical rotation $R$ and 2-fold horizontal magnetic rotation axis $\bsc\equiv\bsm_\text{v}\bsm_\text{h}\bst$, it's defined by the following algebraic relations
\bea
R^n=\bsc^2=(R\cdot\bsc)^2=1.
\eea
where both $\bsc$ and $R\cdot\bsc$ corresponds to in-plane (horizontal) 2-fold magnetic rotation axes. Its projective representation
\bea\label{gpc:2,Cnv,Za}
\gpc{2}{Z_n\rtimes Z_2^\bst}{\mbz_a}=\mbz_{(n,a,2)}\times\mbz_{(n,a)}\times\mbz_{(2,a)}
\eea
is characterized by $\mbz_a$-valued factors
\bea
(U_R)^n=\omega_{C_n}\equiv e^{\frac{2\pi\imth}{a}\nu_{C_n}},\\
U_\bsc U^\ast_\bsc=\omega_{\bsc}= e^{\frac{2\pi\imth}{a}\nu_{\bsc}},\\
U_RU_\bsc(U_RU_\bsc)^\ast=\omega_{R\bsc}= e^{\frac{2\pi\imth}{a}(\nu_{R\bsc}+\nu_\bsc)}.
\eea
It is straightforward to show that the solutions to the factors are
\bea
&\nu_{C_n}\simeq\nu_{C_n}+n\mod a\in\mbz_{(n,a)},\\
&2\nu_\bsc=0\mod a\Longrightarrow\nu_\bsc\in\mbz_{(n,a,2)},\\
&2\nu_{R\bsc}=n\nu_{R\bsc}=0\mod a\Longrightarrow\nu_{R\bsc}\in\mbz_{(n,a,2)}.
\eea
as shown earlier in (\ref{k=2,d=3:magnetic Cnv}). Physically they correspond to the topological index of 1d $G_0$-SPT phases decorated on the $C_n$ rotation axis ($\nu_{C_n}$, colored blue in FIG. \ref{fig:k=2:Cnv}), on each horizontal $\bsc$ axis ($\nu_\bsc$, colored red in FIG. \ref{fig:k=2:Cnv}), and on each horizontal $R\cdot\bsc$ axis ($\nu_{R\bsc}$, colored green in FIG. \ref{fig:k=2:Cnv}), as shown in FIG. \ref{fig:k=2:Cnv}.

Clearly there are robust corner states at each intersection of the surface with the 3 types of rotation axes: vertical $n$-fold rotation $R$, horizontal 2-fold magnetic rotation $\bsc$ and $R\cdot\bsc$. All these corner states are protected by onsite symmetry $G_0$ and are robust against disorders and crystal distortions.

Compared to magnetic point group $C_n\rtimes Z_2^{\bsm_\text{v}\bsm_\text{h}\bst}$, the point group $G_c=C_n$ case is different. In addition to $n$-fold vertical rotation axis $R$, it also has vertical mirror planes $\bsm_\text{v}$ and $R\cdot\bsm_\text{v}$. Although the Kunneth formula (\ref{k=2,d=3:Cnv}) yields the same outcome as the magnetic point group in (\ref{k=2,d=3:magnetic Cnv}), the factors have different meanings. While $\nu_{C_n}$ still labels the topological index of 1d $G_0$-SPT phase decorated on the vertical $C_n$ rotation axis, $\nu_\bsc$ and $\nu_{R\bsc}$ correspond to weak crystalline SPT indices. They characterize whether each 2d mirror plane, $\bsm_\text{v}$ and $R\cdot\bsm_\text{v}$, are 2d SPT phases protected by mirror and $G_0$ symmetries. Although there can be gapless boundary states if mirror symmetry is preserved by the surface, they are generally not stable against perturbations breaking the mirror symmetry on the surface.

%\subsubsection{$G_c=D_{n,\text{h}}$}
%
%\bea
%\gpc{2}{(Z_n\rtimes Z_2)\times Z_2^\bst}{\mbz_a}=\mbz_{(n,a,2)}^2\times\mbz_{(2,a)}^4
%\eea

\begin{figure}
\includegraphics[width=\columnwidth]{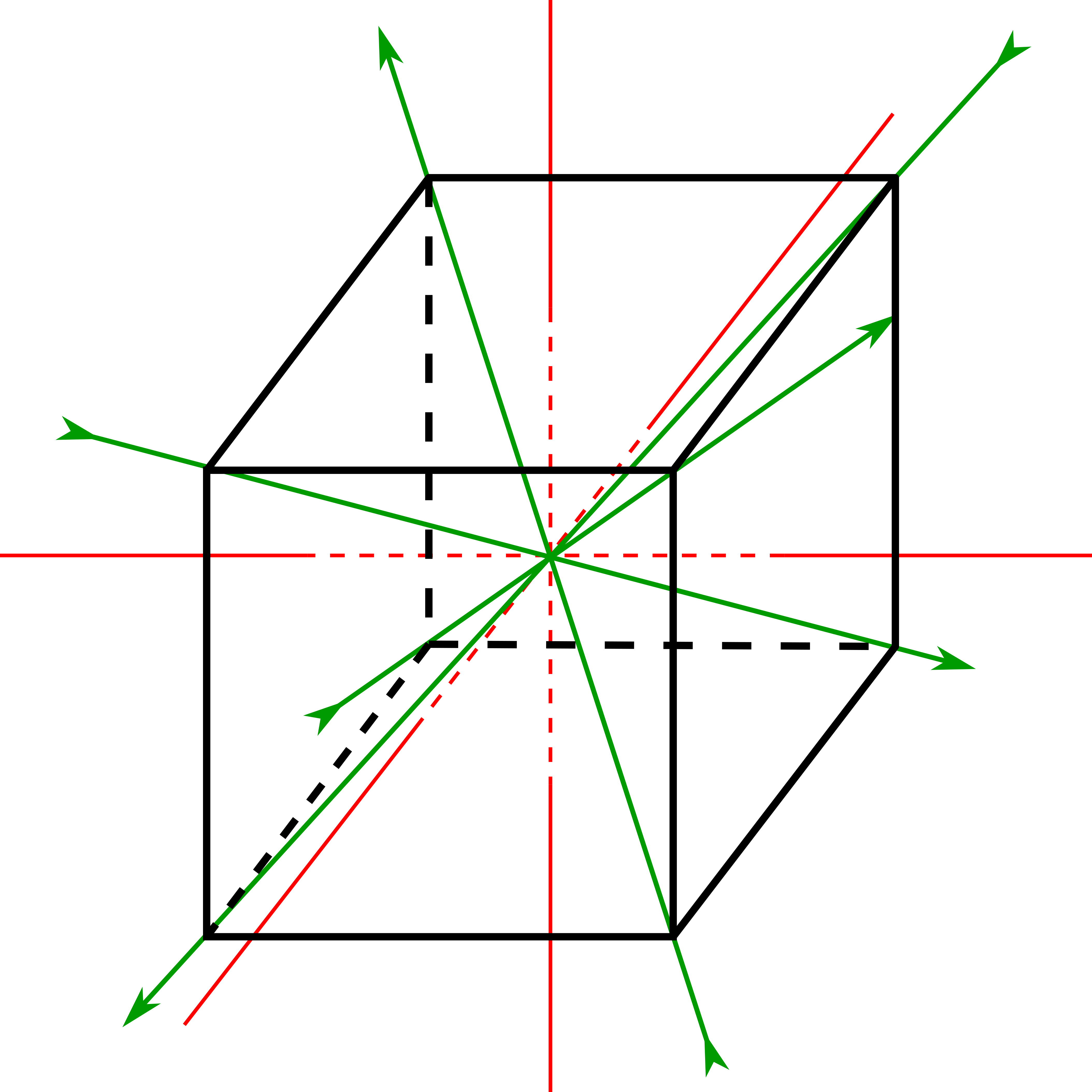}
\caption{3rd-order SPT phases with point group symmetry $T$. They are constructed by decorating the four 3-fold axes (green) with 1d $G_0$-SPT phases with topological index $\nu_3$, and the three 2-fold axes (red) with topological index $\nu_2$.}
\label{fig:T}
\end{figure}

\subsubsection{$G_c=T$}

Point group $T$ is generated by 2-fold rotations $R_{y,z}$ along $\hat y$ and $\hat z$ axes, as well as 3-fold rotations $R_3$ along (111) axis:
\bea
T\equiv\{R_y^{i_y}R_z^{i_z}R_3^{i_3}|i_{y,z}\in Z_2,~i_3\in Z_3\}.
\eea
The group multiplication rules are set by the following algebraic identities:
\bea\label{algebra:T:1}
&(R_y)^2=(R_z)^2=(R_yR_z)^2=(R_3)^2=1,\\
&R_3R_zR_3^{-1}=R_yR_z,~~~R_3R_yR_3^{-1}=R_z.\label{algebra:T:2}
\eea
Its projective representation are determined by the following phase factors
\bea
&(U_{R_y})^2=(U_{R_z})^2=(U_{R_y}U_{R_z})=\omega_{2}=e^{\frac{2\pi\imth}{a}\nu_2}\in\mbz_a,~~~\\
&(U_{R_3})^3=\omega_{3}=e^{\frac{2\pi\imth}{a}\nu_3}\in\mbz_a.
\eea
It's straightforward to show that
\bea
&\nu_3\simeq\nu_3+3\mod a\Longrightarrow\nu_3\in\mbz_{(a,3)},\\
&\nu_2=-\nu_2\mod a\Longrightarrow\nu_2\in\mbz_{(a,2)}.
\eea
leading to the group cohomology classification
\bea
\gpc{2}{T^\ast\simeq A_4}{\mbz_a}=\mbz_{(3,a)}\times\mbz_{(2,a)}
\eea

As shown in FIG. \ref{fig:T}, the $T$-symmetric 3rd-order SPT phases are constructed by decorating all four of the 3-fold axes (green in FIG. \ref{fig:T}) by 1d $G_0$-SPT phases with topological index $\nu_3\in\mbz_{(3,a)}$, and decorating all three of the 2-fold axes (red in FIG. \ref{fig:T}) by 1d $G_0$-SPT phases with topological index $\nu_2\in\mbz_{(2,a)}$.

Now we discuss $G_0$ symmetry protected corner states in the system. If the surface intersects with any of the 2-fold or 3-fold axes, it will host gapless corner modes protected by onsite symmetry $G_0$ at the intersection. Generally the projective representations for corners on the 2-fold and 3-fold axes will be different. Take a spin-1 system with $G_0=SO(3)$ symmetry for example, there will only be gapless spin-1/2 modes at each corner on the $\hat x$, $\hat y$ and $\hat z$ axes, but not on (111) axis.

\begin{figure}
\includegraphics[width=0.8\columnwidth]{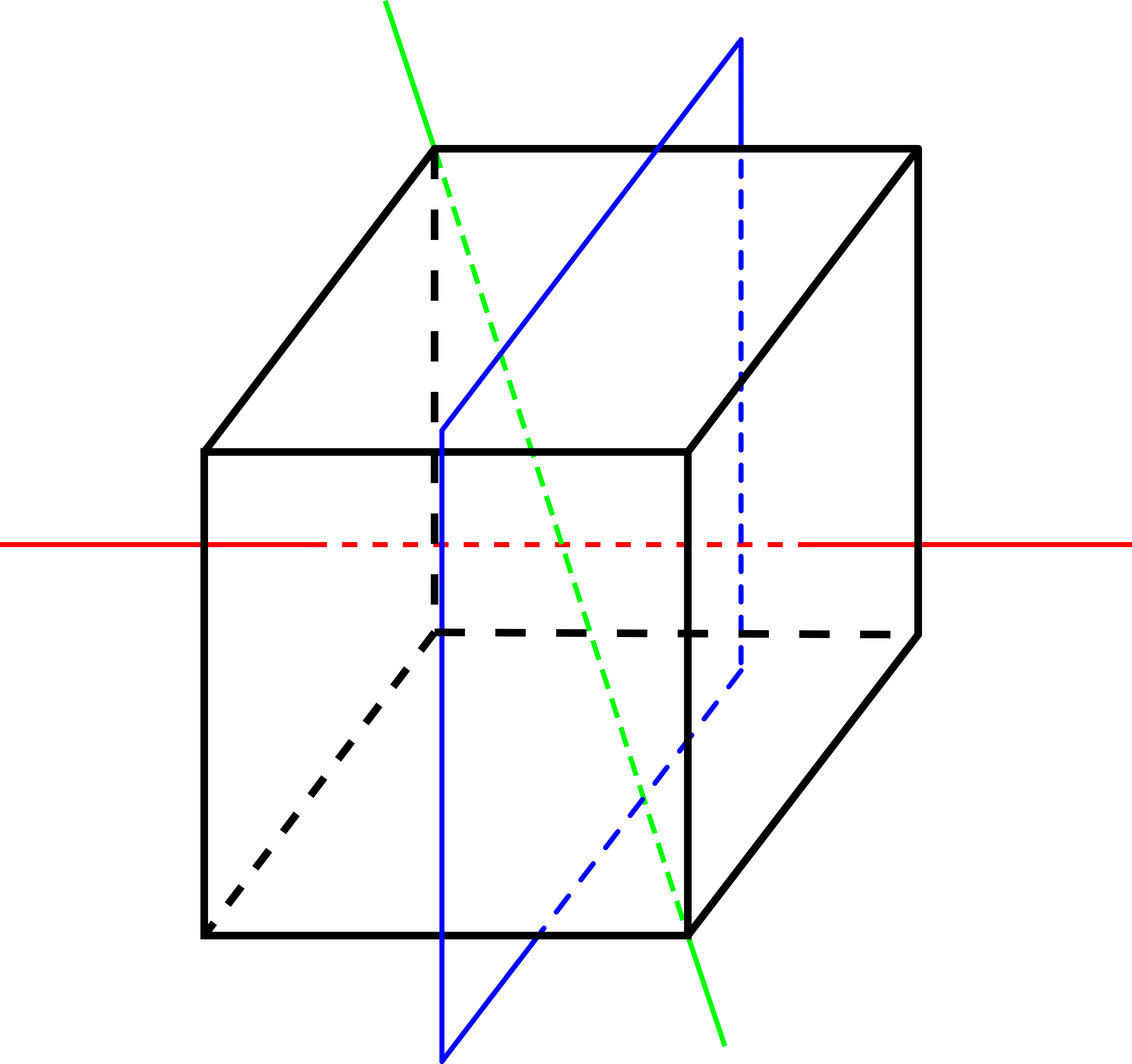}
\caption{3rd-order SPT phases with point group symmetry $T_h$, classified by three topological invariants $\nu_{1,2,3}$. Among them $\nu_{2,3}$ correspond to the topological indices of 1d $G_0$-SPT phases along the 2 types of high-symmetry axes colored by red and green, while $\nu_1+\nu_2$ labels whether each mirror plane (perpendicular to red axis) is a 2d SPT protected by both mirror and onsite symmetry $G_0$.}
\label{fig:Th}
\end{figure}

\subsubsection{$G_c=T_h=T\times Z_2^{\bsi}$}

Finally we consider the point group $T_h$, which is a direct product of group $T$ and the rank-2 group $Z_2^\bsi$ generated by inversion $\bsi$. In addition to algebraic relations (\ref{algebra:T:1})-(\ref{algebra:T:2}), we also have
\bea
&\bsi^2=1,\\
&\bsi R_\alpha\bsi^{-1}=R_\alpha,~~~\alpha=y,z,3.
\eea
This leads to 3 more phase factors, in addition to $\omega_{2,3}$ considered in $G_c=T$ case:
\bea
&U_\bsi U^\ast_\bsi=\omega_1=e^{\frac{2\pi\imth}{a}\nu_1},\\
&U_\bsi U_{R_{y,z}}^\ast U^{-1}_\bsi U_{R_{y,z}}^{-1}=\omega_{4}=e^{\frac{2\pi\imth}{a}\nu_4},\\
&U_\bsi U_{R_3}^\ast U_\bsi^{-1}U_{R_3}^{-1}=\omega_5=e^{\frac{2\pi\imth}{a}\nu_5}.
\eea
It is straightforward to show that $\omega_{4,5}\equiv1$, and
\bea
2\nu_i=0\mod a\Longrightarrow \nu_i\in\mbz_{(2,a)},~~~i=1,2,3.
\eea
This results in the 2nd group cohomology classification
\bea
\gpc{2}{T_\text{h}^\ast\simeq A_4\times Z_2^\bst}{\mbz_a}=\mbz_{(2,a)}^3
\eea
Physically similar to $G_c=T$ case discussed earlier, $\nu_{2,3}$  still correspond to the topological indices of 1d $G_0$-SPT phases, assigned along the 2 types of high-symmetry axes (and their symmetry-related partners) colored by red (with index $\nu_2$) and green (with index $\nu_3$) as shown in FIG. \ref{fig:Th}. As a result, if the surface of an open system intersects with one of these axes, it will host gapless corner modes protected by onsite symmetry $G_0$. A difference between this case and the previous $G_c=T$ example is that due to inversion symmetry, each 1d SPT phase decorated along a high-symmetry axis must be its own inverse phase, leading to $\omega_i=\omega_i^\ast=\pm1$.

Meanwhile $\nu_1$ has a slightly different physical meaning. Notice that there are 3 mirror planes associated with mirror reflection symmetry $\bsm_\alpha=\bsi\cdot R_\alpha$ for $\alpha=x,y,z$. Since each mirror symmetry serves as a onsite $Z_2$ symmetry within its 2d mirror plane, he projective representation of the mirror symmetry
\bea
U_{\bsm_\alpha}U_{\bsm_\alpha}^\ast=U_\bsi U_{R_{\alpha}}^\ast U_\bsi^\ast U_{R_{\alpha}}=\omega_1\omega_2\in\mbz_{(2,a)}.
\eea
corresponds to whether each 2d mirror plane is a SPT phase protected by both $Z_2$ mirror symmetry and onsite symmetry $G_0$. In other words, $\omega_1\omega_2=\pm1$ labels whether the mirror domain wall within each mirror plane is decorated by a 1d $G_0$-SPT phase or not. Therefore $\omega_1\omega_2$ is an index for weak crystalline SPT phases, and generally does not host corner/hinge modes robust against small mirror-breaking perturbations.

\section{Discussions}\label{sec:discussions}

In summary, to understand the HOSPT phases of interacting bosons with robust symmetry protected corner/hinge states, we provide a physical picture based on dimensional reduction analysis and a classification and construction based on the Kunneth formula of group cohomology. These strong HOSPT phases support topological boundary excitations robust against general perturbations such as disorders and crystalline distortions, and should be differentiated from weak crystalline SPT phases whose surface states are protected by crystalline symmetries. Focusing on the case where the total symmetry $G=G_c\times G_0$ is a direct product of crystalline symmetry $G_c$ and onsite symmetry $G_0$, we show that a $(k+1)$-th order SPT phase in $d$ spatial dimensions can be built from $G_0$-SPT phases in $(d-k)$ dimensions, and is fully classified within group cohomology formula $\gpc{k}{G_c^\ast}{\gpc{d+1-k}{G_0}{U(1)}}$. Based on a decorated domain wall picture for this group cohomology formula, we show how to explicitly construct a HOSPT phase using lower-dimensional SPT phases as building blocks.

To conclude, we briefly discuss the limitations of the group cohomology classification (\ref{classification}) of HOSPT phases from Kunneth formula. One implicit assumption for the above classification is that the local Hilbert space always forms a linear representation of the total symmetry group. If we consider the local Hilbert space $\mathcal{S}(\alpha)$ at a high-symmetry Wyckoff position $\alpha$, such as the $C_4$ rotation center in FIG. \ref{fig:C4}, the local Hilbert space should also preserve the ``local symmetry'' $G_c(\alpha)=C_4$ in addition to the onsite symmetry $G_0$. In the group cohomology classification (\ref{classification}), we always require such a local Hilbert space $\mathcal{S}(\alpha)$ to form a linear representation of local symmetry $G_c(\alpha)\times G_0$. In particular, the local crystalline symmetry operations in $G_c(\alpha)$ must commute with all onsite symmetry in $G_0$. Failure of this requirement (\ie projective representations of local symmetry $G_c(\alpha)\times G_0$) may lead to even more interesting consequences, such as Lieb-Schultz-Mattis theorems forbidding a short-ranged-entangled ground state\cite{Huang2017b}, which are beyond the description of formula (\ref{classification}).

One natural direction to expand this work is to go beyond a direct product of onsite and crystalline symmetries, and to consider the HOSPT phases with a generic symmetry group. While the Kunneth formula does not simply apply for a generic symmetry group, the dimensional reduction arguments appear to remain valid. Another interesting direction is to use the same approach to study HOSPT phases of interacting fermions. We leave these for future works.

\acknowledgments

We are indebted to Dominic Else, Ying Ran and Shenghan Jiang for helpful discussions, to Chen Fang for feedbacks on the draft, and especially to Dominic Else for pointing out a mistake in the manuscript. YML also thanks Aspen Center for Physics for hospitality, where part of this work was performed. This work is supported by NSF under award number DMR-1653769 (AR,YML), and in part by NSF grant PHY-1607611 (YML). \\

\emph{Note}: In preparation of this work, we became aware of two independent works which studied the general classification of crystalline SPT phases (with onsite and crystalline symmetries) using spectral sequence: one by Dominic Else and Ryan Thorngren\cite{Else2018}, one by Yang Qi and Chen Fang\cite{Song2018a}. Their works have partial overlaps with this work.

\bibliographystyle{naturemag_no_url}
\bibliography{bibs_career}

\end{document}